\documentclass[11pt]{article}

\usepackage{graphicx}
\usepackage{amssymb,latexsym}

\usepackage{xparse}
\usepackage{xspace}
\usepackage{datetime}
\usepackage{amsmath}
\usepackage[labelsep=colon,labelfont=bf]{caption}
\usepackage{tocloft}
\usepackage{pifont}
\usepackage{cite}
\usepackage{color}
\usepackage{collref}
\usepackage[para]{manyfoot}
\usepackage[bottom]{footmisc}
\usepackage{filecontents}
\usepackage[left=2.5cm,right=2.5cm,top=2.5cm,bottom=3cm]{geometry}
\usepackage[linktocpage]{hyperref}

\numberwithin{equation}{section}

\NewDocumentCommand\eqn{mo}{%
  \IfNoValueTF{#2}
     {\[ #1 \]}
     {\begin{equation}\label{#2} #1 \end{equation} \expandafter\newcommand\csname #2\endcsname{\eqref{#2}\xspace}\ignorespaces}
}
\NewDocumentCommand\eqna{mo}{%
  \IfNoValueTF{#2}
    {\begin{align*} #1 \end{align*}}
    {\begin{equation}\label{#2}\begin{split} #1 \end{split}\end{equation} \expandafter\def\csname #2\endcsname{\eqref{#2}\xspace}\ignorespaces}
}
\NewDocumentCommand\twoseqn{momoo}{%
    \IfNoValueTF{#5}
       {\begin{subequations}\begin{align} #1\label{#2} \\ #3 \label{#4}  \end{align}\end{subequations} \expandafter\def\csname #2\endcsname{\eqref{#2}\xspace}\ignorespaces \expandafter\def\csname #4\endcsname{\eqref{#4}\xspace}\ignorespaces}
       {\begin{subequations}\label{#5}\begin{align} #1\label{#2} \\ #3 \label{#4}  \end{align}\end{subequations} \expandafter\def\csname #5\endcsname{\eqref{#5}\xspace}\ignorespaces \expandafter\def\csname #2\endcsname{\eqref{#2}\xspace}\ignorespaces \expandafter\def\csname #4\endcsname{\eqref{#4}\xspace}\ignorespaces}
}
\NewDocumentCommand\threeseqn{momomoo}{%
   \IfNoValueTF{#7}
     {\begin{subequations}\begin{align} #1\label{#2} \\ #3 \label{#4} \\ #5 \label{#6} \end{align}\end{subequations} \expandafter\def\csname #2\endcsname{\eqref{#2}\xspace}\ignorespaces \expandafter\def\csname #4\endcsname{\eqref{#4}\xspace}\ignorespaces \expandafter\def\csname #6\endcsname{\eqref{#6}\xspace}\ignorespaces}
     {\begin{subequations}\label{#7}\begin{align} #1\label{#2} \\ #3 \label{#4} \\ #5 \label{#6} \end{align}\end{subequations} \expandafter\def\csname #7\endcsname{\eqref{#7}\xspace}\ignorespaces \expandafter\def\csname #2\endcsname{\eqref{#2}\xspace}\ignorespaces \expandafter\def\csname #4\endcsname{\eqref{#4}\xspace}\ignorespaces \expandafter\def\csname #6\endcsname{\eqref{#6}\xspace}\ignorespaces}
}
\NewDocumentCommand\fourseqn{momomomoo}{%
   \IfNoValueTF{#9}
     {\begin{subequations}\begin{align} #1\label{#2} \\ #3 \label{#4} \\ #5 \label{#6} \\ #7\label{#8} \end{align}\end{subequations} \expandafter\def\csname #2\endcsname{\eqref{#2}\xspace}\ignorespaces \expandafter\def\csname #4\endcsname{\eqref{#4}\xspace}\ignorespaces \expandafter\def\csname #6\endcsname{\eqref{#6}\xspace}\ignorespaces \expandafter\def\csname #8\endcsname{\eqref{#8}\xspace}\ignorespaces}
     {\begin{subequations}\label{#9}\begin{align} #1\label{#2} \\ #3 \label{#4} \\ #5 \label{#6} \\ #7\label{#8} \end{align}\end{subequations} \expandafter\def\csname #9\endcsname{\eqref{#9}\xspace}\ignorespaces \expandafter\def\csname #2\endcsname{\eqref{#2}\xspace}\ignorespaces \expandafter\def\csname #4\endcsname{\eqref{#4}\xspace}\ignorespaces \expandafter\def\csname #6\endcsname{\eqref{#6}\xspace}\ignorespaces \expandafter\def\csname #8\endcsname{\eqref{#8}\xspace}\ignorespaces}
}

\NewDocumentCommand\newsec{mo}{%
  \IfNoValueTF{#2}
     {\section{#1}}
     {\section{#1}\label{#2} \expandafter\gdef\csname #2\endcsname{\ref{#2}\xspace}\ignorespaces}
}
\NewDocumentCommand\subsec{mo}{%
  \IfNoValueTF{#2}
     {\subsection{#1}}
     {\subsection{#1}\label{#2}\expandafter\gdef\csname #2\endcsname{\ref{#2}\xspace}\ignorespaces}
}
\NewDocumentCommand\subsubsec{mo}{%
  \IfNoValueTF{#2}
     {\subsubsection{#1}}
     {\subsubsection{#1}\label{#2}\expandafter\gdef\csname #2\endcsname{\ref{#2}\xspace}\ignorespaces}
}

\makeatletter
\renewcommand\section{\@startsection {section}{1}{\z@}%
{-6ex \@plus -1ex \@minus -.2ex}%
{2.3ex \@plus.2ex}%
{\bfseries}}
\makeatother
\makeatletter
\renewcommand\subsection{\@startsection{subsection}{2}{\z@}%
                                     {-3.25ex\@plus -1ex \@minus -.2ex}%
                                     {1.5ex \@plus .2ex}%
                                     {\itshape}}
\makeatother
\makeatletter
\renewcommand\subsubsection{\@startsection{subsubsection}{3}{\z@}%
                                     {-3.25ex\@plus -1ex \@minus -.2ex}%
                                     {1.5ex \@plus .2ex}%
                                     {\itshape}}
\makeatother
\makeatletter 
\def\@seccntformat#1{\csname the#1\endcsname.\hspace{4.6pt}} 
\makeatother 

\renewcommand{\appendix}{\appendices}

\newenvironment{acknowledgments}{\vspace{12pt}\begin{center}\textbf{Acknowledgments}\end{center}\vspace{-12pt}}{}

\collectsep{\ding{70}~}
\newcommand{\foot}{\footnote}
\newdateformat{mydate}{\monthname[\THEMONTH] \THEYEAR} 
\mydate{}

\DeclareNewFootnote[para]{E}[roman]
\newcommand{\email}[1]{\footnoteE{\href{mailto:#1}{\texttt{#1}}}}
\newcommand{\emails}[1]{\let\thefootnote\relax\footnotetext{{\texttt{#1}}}}

\makeatletter
	\renewcommand{\abstract}[1]{\def \@abstract {#1}}
	\newcommand{\affiliation}[1]{\def \@affiliation {#1}}
	\newcommand{\preprint}[1]{\def\@preprint {#1}}
	\abstract{}
	\affiliation{}
	\preprint{}
\makeatother

\makeatletter
\def \maketitle {%
	\begin{titlepage}
	          \begin{flushright}
                           \@preprint
                   \end{flushright}
                           \vspace{2cm}
		\begin{center}
			{\Large\bfseries \@title} 
		
			\bigskip\bigskip\bigskip
		
			\@author 
			
			\bigskip
			
			\emph{\@affiliation}
			
                  	\end{center}
			\bigskip

			\noindent\@abstract
			
			\vfill\vfill\vfill\vfill\vfill\vfill\vfill\vfill\vfill\vfill\vfill
			\vfill\vfill\vfill\vfill\vfill\vfill\vfill\vfill\vfill\vfill\vfill

			\noindent\@date
	\end{titlepage}
}
\makeatother

\usepackage{epsfig}
\usepackage{graphics}
\usepackage{indentfirst}
\usepackage{amsmath, amsthm}

\let\a=\alpha \let\b=\beta \let\g=\gamma \let\d=\delta \let\e=\epsilon
  \let\tt=\theta  \let\k=\kappa
\let\l=\lambda \let\m=\mu \let\n=\nu \let\x=\xi \let\p=\pi 
\let\s=\sigma   \let\f=\phi  
 
\let\w=\omega        \let\Th=\Theta \let\L=\Lambda
\let\X=\Xi  \let\S=\Sigma  \let\Y=\Psi
 
\let\la=\label  
  
 \def\bd{\begin{document}} \def\ed{\end{document}}
\def\ds{\documentstyle} \let\fr=\frac \let\bl=\bigl \let\br=\bigr
\let\Br=\Bigr \let\Bl=\Bigl
\let\bm=\bibitem
\let\na=\nabla
\def\tU{{\widetilde U}}
\let\pa=\partial \let\ov=\overline
\def\ie{{\it i.e.\ }}
\newcommand{\be}{\begin{equation}}
\newcommand{\ee}{\end{equation}}
\def\ba{\begin{array}}
\def\ea{\end{array}}

\def\bei{\begin{itemize}}
\def\eei{\end{itemize}}

\def\ben{\begin{enumerate}}
\def\een{\end{enumerate}}

\def\ft#1#2{{\textstyle{{\scriptstyle #1}\over {\scriptstyle #2}}}}
\def\fft#1#2{{#1 \over #2}}
\def\F#1#2{{ F_{#1}^{(#2)} }}
\def\cF#1#2{{ {\cal F}_{#1}^{(#2)} }}

\def\R{{\bf R}}
\def\sst#1{{\scriptscriptstyle #1}}
\def\oneone{\rlap 1\mkern4mu{\rm l}}
\def\e7{E_{7(+7)}}
\def\td{\tilde}
\def\wtd{\widetilde}
\def\im{{\rm i}}
\def\bog{Bogomol'nyi\ }
\newcommand{\ho}[1]{$\, ^{#1}$}
\newcommand{\hoch}[1]{$\, ^{#1}$}
\newcommand{\bea}{\begin{eqnarray}}
\newcommand{\eea}{\end{eqnarray}}
\newcommand{\ra}{\rightarrow}
\newcommand{\lra}{\longrightarrow}
\newcommand{\Lra}{\Leftrightarrow}
\newcommand{\bp}{\tilde \beta^\prime}
\newcommand{\cB}{{\cal B}}
\newcommand{\cO}{{\cal O}}
\newcommand{\vecx}{\vec{x}}
\newcommand{\vecy}{\vec{y}}
\newcommand{\vecp}{\vec{p}}
\newcommand{\vecq}{\vec{q}}
\newcommand{\tr}{{\rm tr} }
\newcommand{\Tr}{{\rm Tr} }
\newcommand{\NP}{Nucl. Phys. }

\newcommand{\cL}{{\cal L}}
\newcommand{\cA}{{\cal A}}
\newcommand{\cT}{{\cal T}}
\newcommand{\cD}{{\cal D}}
\newcommand{\cH}{{\cal H}}

\def\sst#1{{\scriptscriptstyle #1}}
\def\0{{\sst{(0)}}}
\def\1{{\sst{(1)}}}
\def\2{{\sst{(2)}}}
\def\3{{\sst{(3)}}}
\def\4{{\sst{(4)}}}
\def\5{{\sst{(5)}}}
\def\6{{\sst{(6)}}}
\def\7{{\sst{(7)}}}
\def\8{{\sst{(8)}}}
\def\9{{\sst{(9)}}}
\def\p{{\sst{(p)}}}
\def\q{{\sst{(q)}}}
\def\sx{{\sst{(x)}}}
\def\sy{{\sst{(y)}}}
\def\ve{\varepsilon}
\def\vf{\varphi}
\def\F{\Phi}
\def\wg{\wedge}

\def\thb{\bar{\theta}}
\def\Thb{\bar{\Theta}}
\def\barp{\bar{p}}
\def\barq{\bar{q}}
\def\barc{\bar{c}}
\def\bard{\bar{d}}
\def\e{\epsilon}

\def \bi{\bibitem}
\def \la {\label}

\def \l {\lambda}
\def \tl  {{\tilde \l}}
\def \sql {{\sqrt \l}}
\def \adss {$AdS_5 \times S^5$\ }
\newcommand{\rf}[1]{(\ref{#1})}
\def \ov {\over}

\def\th{\theta}
\def\Th{\Theta}
\def\vth{\vartheta}
\def\btheta{{\bar\theta}}
\def\ttheta{{{\tilde\theta}}}
\def\bttheta{{{\bar\ttheta}}}
\def\vth{\vartheta}

\def\ra{\rightarrow}
\def\N{{\cal N}}
\def\F{{\cal F}}
\def\uM{\underline{M}}
\def\uA{\underline{A}}
\def\uN{\underline{N}}
\def\uP{\underline{P}}
\def\ua{\underline{a}}
\def\ub{\underline{b}}
\def\uc{\underline{c}}
\def\ud{\underline{d}}
\def\ue{\underline{e}}
\def\uf{\underline{f}}
\def\ui{\underline{i}}
\def\uj{\underline{j}}
\def\uk{\underline{k}}
\def\ual{\underline{\alpha}}
\def\ube{\underline{\beta}}
\def\um{\underline{m}}
\def\un{\underline{n}}
\def\up{\underline{p}}
\def\uq{\underline{q}}
\def\ur{\underline{r}}
\def\us{\underline{s}}

\def\umu{\underline{\mu}}
\def\unu{\underline{\nu}}
\def\ula{\underline{\l}}
\def\uka{\underline{\k}}
\def\usi{\underline{\s}}
\def\urh{\underline{\r}}
\def\cc{\circ}
\def\eqv{\equiv}

\def\ni{\noindent}

\def\Ep{E^{{}^{(+)}}}
\def\Em{E^{{}^{(-)}}}

\def\Mp{M^{{}^{(+)}}}
\def\Mm{M^{{}^{(-)}}}

\def \ha{{1\ov 2}}

\def\r{\rho}

\def\Y{{\rm Y}}
\def\X{{\rm X}}
\def\tY{\tilde{\rm Y}}
\def\tX{\tilde{\rm X}}
\def\dY{\dot{\rm Y}}
\def\dX{\dot{\rm X}}

\def \J {\mathcal{J}}
\def \del {\partial}

\def\dF{\dot{F}}
\def\dG{\dot{G}}
\def\dx{\dot{x}}
\def\de{\dot{e}}
\def\dr{\dot{r}}
\def\dt{\dot{t}}
\def\dth{\dot{\tt}}
\def\df{\dot{\phi}}

\def\ddx{\ddot{x}}
\def\ddt{\ddot{t}}
\def\ddr{\ddot{r}}
\def\ddth{\ddot{\tt}}
\def\ddf{\ddot{\phi}}

\def \E {{\cal E}}
\def \S {{\cal S}}
\def \J {{\cal J}}

\def\ms{\mathcal{S}}
\def\mj{\mathcal{J}}
\def\soj{\fr{\ms}{\mj}}
\def \R {{\bf R}}
\def \om {\omega}
\def \bE {\bar E}
\def \x {{\cal X}}

\def \bi{\bibitem}
\def \la {\label}

\def \l {\lambda}
\def\foot{\footnote}
\def \tl  {{\tilde \l}}
\def \sql {{\sqrt \l}}
\def \adss {$AdS_5 \times S^5$\ }
\def \ov {\over}

\def \varpi {{\rm w}}

\def\thb{\bar{\theta}}
\def\Thb{\bar{\Theta}}
\def\mb{\bar{\m}}
\def\ab{\bar{\a}}
\def\zb{\bar{z}}
\def\psib{\bar{\psi}}
\def\barp{\bar{p}}
\def\barq{\bar{q}}
\def\barc{\bar{c}}
\def\bard{\bar{d}}
\def\e{\epsilon}
\def\wb{\bar{w}}
\def\lb{\bar{\l}}
\def\Jb{\bar{J}}
\def\Nb{\bar{N}}
\def\Zb{\bar{Z}}
\def\pab{\bar{\pa}}

\def\bg{\bar{g}}

\def\At{\tilde{A}}
\def\Bt{\tilde{B}}
\def\Ct{\tilde{C}}
\def\Dt{\tilde{D}}
\def\Et{\tilde{E}}
\def\Ft{\tilde{F}}
\def\Gt{\tilde{G}}
\def\Ht{\tilde{H}}
\def\It{\tilde{I}}
\def\Mt{\tilde{M}}
\def\Rt{\tilde{R}}
\def\St{\tilde{S}}
\def\at{\tilde{a}}
\def\bt{\tilde{b}}
\def\ct{\tilde{c}}
\def\et{\tilde{e}}
\def\ft{\tilde{f}}
\def\gt{\tilde{g}}
\def\mt{\tilde{\mu}}
\def\nt{\tilde{\nu}}

\def\asth{\hat{*}}
\def\phh{\hat{\phi}}

\def\bA{{\bf A}}

\def\ola{\overleftarrow}
\def\ora{\overrightarrow}
\def\alt{\tilde{\a}}

\def\ra{\rightarrow}
\def\Ra{\Rightarrow}

\def\eh{\hat{e}}
\def\eph{\hat{\e}}
\def\ph{\hat{p}}
\def\alh{\hat{\a}}
\def\beh{\hat{\b}}
\def\gah{\hat{\g}}
\def\Fh{\hat{F}}
\def\muh{\hat{\m}}
\def\nuh{\hat{\n}}
\def\thh{\hat{\th}}
\def\dh{\hat{d}}
\def\ih{\hat{i}}
\def\jh{\hat{j}}
\def\kh{\hat{k}}
\def\deh{\hat{\d}}
\def\wh{\hat{w}}
\def\lah{\hat{\l}}
\def\Ah{\hat{A}}
\def\Ch{\hat{C}}
\def\Omh{\hat{\Omega}}

\def\xh{\hat{x}}

\def\ps{\rlap{\, /}\;\,p }
\def\ks{\rlap{\, /}\;\,k }

\def\gym{g_{YM}}

\def\adot{\dot{a}}
\def\bdot{\dot{b}}
\def\bpa{\bar{\pa}}

\def\pr{\prime}
\def\ssk{\medskip}
\def\bsk{\bigskip}
\def\N{\nabla}
\def\clb{\color{blue}}
\def\clr{\color{red}}
\def\clv{\colo{violet}}

\def\lras{\leftrightarrows}

\def\mx{\mathcal{X}}
\def\my{\mathcal{Y}}

\title{
Liouville mode in gauge/gravity duality
}

\author{Tatiana Moskalets$\,^{\spadesuit,}$\email{tatyana.moskalets@gmail.com}, Alexei Nurmagambetov$\,^{\diamondsuit,\spadesuit,}$\email{ajn@kipt.kharkov.ua}}

\affiliation{
$^{\spadesuit}$ Department of Physics and Technology, Karazin Kharkov National University, 4 Svobody Sq., Kharkov, UA 61022
Ukraine\\

$^{\diamondsuit}$
Akhiezer Institute for Theoretical Physics of
NSC KIPT,\\
1 Akademicheskaya St., Kharkov, UA 61108 Ukraine
}

\abstract{

We establish solutions corresponding to AdS$_4$ static charged black holes with inhomogeneous two-dimensional horizon surfaces of constant curvature. Depending on the choice of the 2D constant curvature space, the metric potential of the internal geometry of the horizon satisfies the elliptic wave/elliptic Liouville equations.
We calculate the charge diffusion and transport coefficients in the hydrodynamic limit of gauge/gravity duality and observe the exponential suppression in the diffusion coefficient and in the shear viscosity-per-entropy density ratio in the presence of an inhomogeneity on black hole horizons with planar, spherical, and hyperbolic geometry. We discuss the subtleties of the approach developed for a planar black hole with inhomogeneity distribution on the horizon surface in more detail and find, among others, a trial distribution function, which generates values of the shear viscosity-per-entropy density ratio falling within the experimentally relevant range. The solutions obtained are also extended to higher-dimensional AdS space. We observe two different DC conductivities in 4D and higher-dimensional effective strongly coupled dual media and formulate conditions under which the appropriate ratio of different conductivities is qualitatively the same as that observed in an anisotropic strongly coupled fluid. We briefly discuss ways of how the Liouville field could appear in condensed matter physics and outline prospects of further employing the gauge/gravity duality in CMP problems.

\ssk\ssk
{PACS numbers: 04.70.-s, 05.60.-k, 04.40.Nr, 04.20.Jb, ok}

}

\date{\today}

\begin{document}
\maketitle

\tableofcontents

\section{Introduction}

This paper is motivated by recent progress in applying the AdS/CFT correspondence to condensed matter physics. Since several  universal bounds (momentum $\eta/s \ge 1/4\pi$ and charge $\s_{DC}/\chi \ge d/4\pi T (d-2)$ transport bound relations in holographic hydrodynamics \cite{Kovtun:2003wp, Kovtun:2004de, Kovtun:2008kx, Ritz:2010zza}, $\w_g/T_c \gtrsim 8$ in holographic super\-con\-duc\-tivity \cite{Horowitz:2008bn}) have been established for strongly coupled effective dual media, it is reasonable to pose the question: How robust are these relations? If one is limited to the standard gravitational theory setup no signs of violation of these universal bounds have been found.\footnote{Within the assumptions made on the structure of the bulk metric (see, e.g., \cite{Iqbal:2008by}).}
However, different extensions of general relativity with higher order curvature terms revealed the violation of these universal relations (see, e.g., \cite{Brigante:2007nu, Ritz:2008kh, Wu:2010vr}).
But still the question remains: may violations of the universal bounds of gauge/gravity duality be found within the Einstein theory?
In fact, the positive answer to this question is known and it is related to introducing {\it anisotropy} \cite{Erdmenger:2010xm, Mateos:2011ix, Mateos:2011tv, Rebhan:2011vd, Critelli:2014kra} on the horizon surface in black hole (BH) solutions.

\ssk
Indeed, the origin of universality in gauge/gravity duality is closely related to properties of black holes. In further discussion two observations will be important:
\begin{enumerate}
\item
Thermodynamics of static charged black holes is fully managed by the $(g_{tt},g_{rr})$ parts of the metric\foot{For the metric structure encoded in the space-time interval $\mathrm{d}s^2=g_{tt}(r)\mathrm{d}t^2+g_{rr}(r)\mathrm{d}r^2+r^2\gamma_{ij} \mathrm{d}X^i \mathrm{d}X^j$ with coordinates $t,r,X,Y,\dots$ and diagonal internal metric $\g_{ij}$.} and depends on global geometry of the horizon surface, more precisely on its integral volume.
\item
The transport coefficients are determined by the local geometry of the horizon surface, as well as by the $g_{rr}$ part of the metric, which determines the radial coordinate value of the horizon location.
\end{enumerate}

\ssk
On account of these facts it is easy to see that either the modifications in the $(g_{tt},g_{rr})$ parts of BH solutions due to the change of the bulk gravitational dynamics \cite{Brigante:2007nu, Wu:2010vr, Ritz:2008kh} (without changing the horizon geometry) or changing the horizon surface geometry to the anisotropic one \cite{Erdmenger:2010xm, Mateos:2011ix, Mateos:2011tv, Rebhan:2011vd, Critelli:2014kra} (without changing the standard dynamics of the bulk gravity) should lead to corrections to universal relations. Note, however, that the latter case requires introducing additional fields, as compared to the standard Einstein--Maxwell system.

\ssk
In this paper we will limit ourselves with the standard dynamics managed by the Einstein--Hilbert--Maxwell action and will take a look at universality in gauge/gravity duality from a different angle. It is clear from the discussion above that the universality violation will require changing in the horizon geometry. In \cite{Erdmenger:2010xm, Mateos:2011ix, Mateos:2011tv, Rebhan:2011vd, Critelli:2014kra} the standard geometry of the horizon surface was changed to the anisotropic one, modifying the planar geometry of the horizon with factoring one of the horizon coordinates by a function of the radial coordinate $\mathcal{H}(r)$. By means of differential geometry it is easy to see that since such a deformation of the horizon is not isometric (it preserves the orthogonality of the coordinate system, but it does not preserve the volume of the horizon), it does not change the geometry type: the external curvature of the horizon surface is still equal to zero. Therefore, one may apply the same computational scheme to calculate the transport coefficients (the AC/DC conductivity and the shear viscosity \cite{Erdmenger:2010xm, Rebhan:2011vd, Critelli:2014kra}) as was done before \cite{Kovtun:2003wp, Kovtun:2004de, Iqbal:2008by}. Our proposal consists in performing another deformation of the horizon surface, which is also not isometric but still keeps the planar geometry: the horizon surfaces considered here are conformally flat. Unlike the deformation previously considered in \cite{Erdmenger:2010xm, Mateos:2011ix}, we consider ``inhomogeneity'' on the horizon surface encoded in function(s) solely dependent on the horizon surface coordinates. But we do it in a way which realises the conformal flatness of the horizon that in its turn justifies employing the technique of \cite{Kovtun:2003wp, Kovtun:2004de} in computing transport coefficients. 

\ssk
The rest of the paper is organised as follows. In Sect. 2 we formulate out setup, which is a standard one in searching for the solutions corresponding to charged black holes in 4D AdS space-time. Then, due to the two-dimensional surface theory, we can write down the part of the metric ansatz, corresponding to geometry of the horizon surface, in terms of isothermal coordinates. Recall that, written in the isothermal coordinates, any two-dimensional surface of genus zero possesses the geometry of conformally flat space. The log of the conformal factor in exponential parameterisation (the so-called metric potential) solely depends on the horizon surface coordinates. In the case of 2D constant curvature horizons the metric potential satisfies the elliptic Liouville equation. This is the way how the {\it Liouville mode}  appears. We establish a general form of the Raissner--N\"ordstrom black hole solutions with electric and magnetic charges, the horizons of which have the planar, spherical and hyperbolic geometry. The Liouville mode of the solution carries the information on inhomogeneity on the horizon surface, which from the point of view of effective dual theory on the boundary of the AdS space plays the role of an inhomogeneity distribution function in a dual medium.

\ssk
In Sect. 3 we calculate the charge diffusion coefficient and the DC conductivity on the inhomogeneous planar horizon of the electrically charged black hole within the stretched horizon approach \cite{Parikh:1997ma} and the hydrodynamic limit of the AdS/CFT correspondence \cite{Kovtun:2003wp, Kovtun:2004de}. Here we derive the Fick law of diffusion in inhomogeneous media and observe the exponential suppression in the diffusion coefficient. The latter results in violation of the universal bound of \cite{Kovtun:2008kx}
for suitable configurations of the metric potential. In Sect. 4 we extend our computations to the shear viscosity in the effective inhomogeneous fluid and establish the same bound value of $\eta/s$ ratio as in \cite{Kovtun:2003wp,Kovtun:2004de}. However, fulfilment of this relation in the considered case reveals the exponential suppression of the KSS \cite{Kovtun:2003wp, Kovtun:2004de} bound value $\eta/s_0=1/4\pi$, computed for the trivial Liouville mode. 

\ssk
Section 5 contains our comments on the transport coefficients in the background of black holes with non-planar inhomogeneous horizons, on occurrence of the Liouville equation in models of condensed matter physics, and on generalisation of the obtained solutions to higher-dimensional AdS spaces. In the latter case we observe two different conductivities on an inhomogeneous horizon of a 5D AdS electrically charged black hole. We establish conditions under which the ratio of different conductivities corresponding to \cite{Rebhan:2011vd} behaves qualitatively the same as 
for the strongly coupled anisotropic plasma model of \cite{Mateos:2011ix, Mateos:2011tv}. In this section we also give an example of the Liouville mode configuration in the planar black hole solution, which preserves the KSS universal bound and fits the experimentally observed upper bound value of $\eta/s$ ratio. 

\ssk
In the last section we present a summary of the results. For the reader's convenience, we add appendices containing the notation and useful information on solutions to the elliptic Liouville equation.

\section{Raissner--N\"ordstrom black holes with inhomogeneity on the horizon surface}

\subsection{Setup}

Let us consider the Reissner--N\"ordstrom (RN)-type solution to the Einstein--Maxwell system in AdS space-time with cosmological constant $\L$, the dynamics of which is described by the following action ($k^2=8\pi G$): 
\bea
\begin{aligned}
{\hat{I}}= & \frac{1}{2k^2}\int_{\mathcal {M}}\mathrm{d}^4 x \sqrt{-g} \,(R-2\L)-\fr14\int_{\mathcal{M}}\mathrm{d}^4
x \sqrt{-g} \,F_{mn}F^{mn} \\
&+\frac{1}{k^2}  \int_{\pa \mathcal{M}}\,\mathrm{d}^3 x \, \sqrt{-h}\,K +\int_{\pa M}\, \mathrm{d}^3x\, \sqrt{-h}\,\,n_m F^{mn}\,  A_n.
\end{aligned}
\la{RNAdSac}
\eea
Integration over $\mathcal{M}$/$\pa \mathcal{M}$ stands for the integration over the AdS/boundary manifold; $h_{mn}$ is the induced metric on the boundary in the $r$ space-time foliation, $n_m=\sqrt{g_{rr}}\d_{mr}$ is the outward normal to the boundary surface.

We will find the RN-type solution to the Einstein--Maxwell equations of motion
\be
\N_m F^{mn} \equiv \fr1{\sqrt{-g}}\,\pa_m\left( \sqrt{-g}\,F^{mn} \right)
=0,
\la{Meom}
\ee
\be
R_{mn}-\frac{1}{2} g_{mn}R-k^2\left( F_{lm}{F^l}_{n}-\frac{1}{4} g_{mn}F_{pq}F^{pq}\right) =-\Lambda g_{mn}
\la{EinEM}
\ee
over the AdS$_4$ background with inhomogeneous horizon metric
\be
 \mathrm{d}s^2=-f(r)\mathrm{d}t^2+\frac{\mathrm{d}r^2}{f(r)}+r^2 (f_1({x},y)
 \mathrm{d}x^2+f_2(x,y) \mathrm{d}y^2),
\la{dsPHgen}
\ee
which generalises the planar (AdS) BH solution \cite{Vanzo:1997gw,Lemos:1994fn}. As usual $\L=-3/l^2$, where $l$ is a characteristic length of the AdS space.

\ssk
\subsection{Solutions for charged AdS BHs}

To complete the task let us recall geometric properties of two-dimensional surfaces, which lie at the basis of string theory. The line element of {\it any} two-dimensional surface of genus zero can be presented in the following way:
\be
\mathrm{d}s^2_{\mathcal{{M}}_2}=\mathrm{e}^{\Phi (x,y)}(\mathrm{d}x^2+\mathrm{d}y^2),
\la{ds22gen}
\ee
where $(x,y)$ are the {\it isothermal} coordinates (see, e.g., Chapter 9, Addendum I of \cite{SpivakV4}). A function $\Phi(x,y)$ is often called the potential of the metric.

\ssk
On account of \rf{ds22gen} we get the following BH solutions to the AdS$_4$ Einstein equation \rf{EinEM} with zero Maxwell field and constant curvature horizon manifolds\footnote{See Appendix A for details on the used notation.}:
\bei
\item
For a planar-type BH
\be
\begin{aligned}&\mathrm{d}s_{{\scriptscriptstyle (0)}}^2=-
 f(r)\mathrm{d}t^2+\frac{\mathrm{d}r^2}{f(r)}+r^2 \mathrm{e}^{\Phi(x,y)}(\mathrm{d}x^2+\mathrm{d}y^2), & f(r)=\left(\fr{r^2}{l^2}-\fr{\w_2 M}{r} \right)
\end{aligned}
\la{dspliso}
\ee
with $\Phi(x,y)$ satisfying the elliptic wave equation
\be
\fr{\pa^2 \Phi}{\pa x^2}+\fr{\pa^2 \Phi}{\pa y^2}=0.
\la{rhoeqpl}
\ee
\item
For a spherical type BH
\be
\begin{aligned}&\mathrm{d}s_{{\scriptscriptstyle (0)}}^2=-
f(r)\mathrm{d}t^2+\frac{\mathrm{d}r^2}{f(r)}+r^2 \mathrm{e}^{\Phi(x,y)}(\mathrm{d}x^2+\mathrm{d}y^2), &f(r)=\frac{r^2}{l^2}-\frac{\omega_2 M}{r} +1
\end{aligned}
\la{dssphiso}
\ee
with $\Phi(x,y)$ satisfying the elliptic Liouville equation
\be
\fr{\pa^2 \Phi}{\pa x^2}+\fr{\pa^2 \Phi}{\pa y^2}+2\mathrm{e}^{\Phi(x,y)}=0.
\la{rhoeqsph}
\ee
\item
For a hyperbolic type BH\footnote{In Schwarzschild coordinates the corresponding interval looks like
\[
\mathrm{d}s^2_{{\scriptscriptstyle (0)}}=-f(r)\mathrm{d}t^2+\frac{\mathrm{d}r^2}{f(r)}+r^2(\mathrm{d}\theta^2+\sinh^2 \tt \,\mathrm{d}\f^2),\qquad f(r)=\fr{r^2}{l^2}-\fr{\w_2 M}{r} -1.
\]
}
\bea
\begin{aligned}
&\mathrm{d}s_{{\scriptscriptstyle (0)}}^2=-f(r)\mathrm{d}t^2+\frac{\mathrm{d}r^2}{f(r)}+r^2 \mathrm{e}^{\Phi(x,y)}(\mathrm{d}x^2+\mathrm{d}y^2),
&f(r)=\frac{r^2}{l^2}-\frac{\omega_2 M}{r} -1
\end{aligned}
\la{dshypiso}
\eea
with $\Phi(x,y)$ satisfying the elliptic Liouville equation
\be
\fr{\pa^2 \Phi}{\pa x^2}+\fr{\pa^2 \Phi}{\pa y^2}-2\mathrm{e}^{\Phi(x,y)}=0.
\la{rhoeqhyp}
\ee
\eei

Recall that the Liouville equation
\be
\fr{\pa^2 \Phi}{\pa x^2}+\fr{\pa^2 \Phi}{\pa y^2}+2K\mathrm{e}^{\Phi(x,y)}=0
\la{LeqK}
\ee
includes the Gauss curvature of 2D manifold $K$. Clearly, the solution \rf{dspliso}--\rf{rhoeqpl} corresponds to the horizon of $K=0$, while the solutions \rf{dssphiso}--\rf{rhoeqsph} and \rf{dshypiso}--\rf{rhoeqhyp} correspond to the horizon surfaces with $K=1$ and $K=-1$, i.e., to a 2D sphere and to a 2D hyperboloid. (See Appendix B for more details as regards physically relevant solutions to the elliptic Liouville equation.)

\ssk
Restoring the Maxwell field we get the following (electrically and magnetically) charged BH solutions to the Einstein--Maxwell system \rf{Meom}--\rf{EinEM}:
\be
\begin{aligned}&\mathrm{d}s^2=-f(r)\mathrm{d}t^2+\frac{\mathrm{d}r^2}{f(r)}+r^2 \mathrm{e}^{\Phi (x,y)}(\mathrm{d}x^2+\mathrm{d}y^2), &
f(r)=\frac{r^2}{l^2}-\frac{\omega_2 M}{r}+K+\sum_{i=e,m}\,\frac{k^2 Q^2_{i}}{2r^2},
\end{aligned}
\la{RNsolem}
\ee
\be
A_m=(A_t(r),0,A_x(x,y),0),\quad A_t(r)=\m-\fr{Q_e}{r},\quad A_x(x,y)=-Q_m \int \,\sqrt{\g_{xx}\,\g_{yy}}\,\mathrm{d}y.
\la{RNAsolem}
\ee
Depending on the Gauss curvature of the horizon surface we choose $K=0,\pm 1$; $Q_{e,m}$ are the values of the electric/magnetic charge densities. As for the neutral BHs, the metric potential $\Phi(x,y)$ has to satisfy the elliptic Liouville equation \rf{LeqK}. The Bianchi identities 
\be
\e^{mnkl} \N_n F_{kl}=0
\la{BIA}
\ee
hold on the vector field ansatz \rf{RNAsolem}.

\ssk
Having established the solutions for the charged AdS BHs with inhomogeneous horizons, let us turn to computations of the related charge/momentum transport coefficients. From now on we will focus on the solution for electrically charged BH with the planar-type inhomogeneous horizon (i.e., on the solution \rf{LeqK}--\rf{RNAsolem} with $K=0$ and $Q_m=0$).

\section{Charge diffusion and DC conductivity }

\subsection{Charge diffusion on a stretched horizon}

A quick way to obtain the charge diffusion coefficient is to use the stretched horizon approach of \cite{Parikh:1997ma} (see also \cite{Blandford:1977ds,Damour:1978cg, Znajek:1978} for early papers on electrodynamics of BHs and \cite{MPbook} and references therein for an introduction to the membrane paradigm approach). 

\ssk
It is a well-known fact (see, e.g., \cite{Iqbal:2008by, Parikh:1997ma}) that the variation of the Maxwell action
\be
S=-\fr14 \int_{r>r_+}\,\mathrm{d}^4x\,\sqrt{-\bg}\,F_{mn}F^{mn},\quad F_{mn}=2\pa_{[m} A_{n]}
\la{Mac}
\ee
in a BH gravitational background field $\bg_{mn}$ leads to the boundary term at the BH horizon\footnote{The horizon is located at $r=r_+$, where $r_+$ is the highest root of equation $f(r)=0$.} 
$r_+$, which is compensated for by the following surface term added to the action (cf. the action \rf{RNAdSac}):
\be
S_{\mathrm{surf}}=\int_{r_\e}\, \mathrm{d}^3 x\, \sqrt{-\bar{h}} \,n_m F^{mn}\,  A_n=\int_{r_\e}\, \mathrm{d}^3 x\, A_m j^m.
\la{Msurf}
\ee
Here $\bar{h}$ is the induced background metric on the stretched horizon $r_\e=r_++\e,~\e \ll 1$ in the $r$ space-time foliation, $n_m=\sqrt{\bg_{rr}}\d_{mr}$ is the outward normal to the horizon surface, and $j^m$ is the conserved current induced on the horizon\footnote{Indeed,
\[
\pa_m j^m=\pa_m\left(\sqrt{-\bar{h}}\,n_n F^{nm}\right)=\sqrt{-\bar{h}}\, F^{nm}\pa_m n_n+n_n \pa_m\left(\sqrt{-\bar{h}}\, F^{nm}\right)=0
\]
by use of the definition of $n_m$ and equation of motion \rf{Meom} ($\sqrt{-\bg}=\sqrt{-\bar{h}}\sqrt{\bg_{rr}}$).}
\be
j^m=\sqrt{-\bar{h}}\,n_n F^{nm}\,\vert_{r_\e},\quad n_m j^m=0,\quad \pa_m j^m=0.
\la{jhor}
\ee

From \rf{jhor} we derive
\refstepcounter{equation}\label{j0i}
\[
j^t=-\sqrt{-\bg}\,\bg^{tt}\,\bg^{rr}\,F_{tr}\,\vert_{r_\e},
\tag{\theequation a}\label{j0}
\]
\[
j^i=\sqrt{-\bg}\,\bg^{rr}\,\bg^{ij}\,F_{rj}\,\vert_{r_\e},\quad i,j=x,y.
\tag{\theequation b}\label{ji}
\]

\ssk
Now we will treat the Maxwell field as a small perturbation over the gravitational back\-gro\-und. In the linear order of perturbations the Einstein--Maxwell system \rf{Meom}--\rf{EinEM} reduces to the AdS$_4$ Einstein equation $R_{mn}(\bg)=\L \bg_{mn}$, which is solved with \rf{dspliso}--\rf{rhoeqpl}, and to the Maxwell field equation of motion $\pa_m(\sqrt{-\bg} F^{mn})=0$ in the background of $\bg_{mn}$. 

\ssk
The conformally flat structure of the horizon geometry makes possible to use the plane wave representation (see Section 3.7. in \cite{BDbook}) for the perturbed Maxwell field. 
Without loss of generality we can choose  
\be
\d A_m=a_m(t,r)\mathrm{e}^{iqx},\quad a_m(t,r) \ll 1.
\la{dAans}
\ee
Calculations of the DC conductivity are performed at $q \ra 0$ (see \cite{Kovtun:2003wp, Iqbal:2008by}).

\ssk
The requirement of the Maxwell field regularity near the horizon imposes the following boundary
condition \cite{Iqbal:2008by}:
\be
F_{rx}=\sqrt{-\fr{\bg_{rr}}{\bg_{tt}}}\,\Bigg|_{r_\e}\, F_{tx}.
\la{Fbc}
\ee
Alternatively, this boundary condition follows from the solution to the Maxwell equations in the near-horizon limit  \cite{Kovtun:2003wp}. Other assumptions lying in the basis of the charge diffusion law derivation and compatible with $q^2/T^2 \ll 1$ vector field series expansion (here $T$ is the BH temperature) are \cite{Kovtun:2003wp} 
\be
\fr{|\pa_t \d A_x|}{|\pa_x \d A_t|} \ll 1 
\la{assumpA1}
\ee
and 
\be
\d A^\0_t(t,r,x)=C_0(t)\,\mathrm{e}^{iqx} \int_r^{\infty}\,\mathrm{d}r'\,\fr{\bg_{tt}(r')\bg_{rr}(r')}{\sqrt{-\bg(r')}}.
\la{assumpA2}
\ee
Adapting the computational scheme of \cite{Kovtun:2003wp} to the considered case one may check the validity of \rf{assumpA1} and \rf{assumpA2}.

\ssk
From \rf{assumpA2} we get \cite{Kovtun:2003wp} 
\be
\fr{\d A_t}{F_{tr}}\Big|_{r_\e}=\sqrt{-\bg (r_+)}\,\bg^{tt}(r_+)\,\bg^{rr}(r_+)\int_{r_+}^\infty\,\mathrm{d}r\,\fr{\bg_{tt}(r)\bg_{rr}(r)}{\sqrt{-\bg (r)}},
\la{AtoFhor}
\ee
and on account of \rf{Fbc} and \rf{assumpA1} we derive
\[\begin{aligned}
j^x= & {} \sqrt{-\bar{g}}\,\bar{g}^{rr}\,\bar{g}^{xx}\,F_{rx}\,\vert_{r_\epsilon }= \sqrt{-
\bar{g}}\sqrt{-\bar{g}^{tt}\bar{g}^{rr}}\,\bar{g}^{xx}\,F_{tx}\,\vert_{r_\epsilon }\\= & {} -\sqrt{-
\bar{g}}\sqrt{-\bar{g}^{tt}\bar{g}^{rr}}\,\bar{g}^{xx}\,\partial_x \delta A_t \,\vert_{r_\epsilon }\\= &
{} -\left( \frac{\delta A_t}{F_{tr}}\right) \sqrt{-\bar{g}}\, \sqrt{-
\bar{g}^{tt}\bar{g}^{rr}}\,\bar{g}^{xx}\,\partial_x F_{tr}\,\vert_{r_\epsilon }.
\end{aligned}\]
Using the definition of \rf{j0} we further get
\[\begin{aligned}
j^x= & {} \left. \left( \frac{\delta A_t}{F_{tr}}\right) \sqrt{-\bar{g}}\,\sqrt{- \bar{g}^{tt}\bar{g}^{rr}}\,\bar{g}^{xx}\,\partial_x \left( \frac{1}{\sqrt{-\bar{g}}\, \bar{g}^{rr}}\,j_t \right)\,\right|_{r_\epsilon }\\= & {} {-}\left. \left( \frac{\delta A_t}{F_{tr}}\right) \frac{\sqrt{-\bar{g}_{tt}\,
\bar{g}_{rr}}}{\bar{g}_{xx}}\,\bar{g}^{tt}\left( \partial_x{-}\frac{1}{\sqrt{-\bar{g}}} \partial_x(\sqrt{-
\bar{g}})\right) j_t\,\right|_{r_\epsilon }.
\end{aligned}\]
Finally, we obtain
\be
j^x=-D \N_x j^t,
\la{Fick1}
\ee
which is the covariantisation of Fick's first law for inhomogeneous media (compare to the corresponding expression in \cite{Kovtun:2003wp}). The diffusion coefficient entering \rf{Fick1} becomes a function of the horizon coordinates $x,y$
\be
D(x,y)=-\fr{\sqrt{-\bg}}{\bg_{xx} \sqrt{-\bg_{tt}\,\bg_{rr}}}\,\Big|_{r_+} \int_{r_+}^\infty\,\mathrm{d}r\,\fr{\bg_{tt}(r)\bg_{rr}(r)}{\sqrt{-\bg (r,x,y)}},
\la{D}
\ee
where we have explicitly marked out the part depending on $x,y$ coordinates. Fick's second law comes from the current conservation:
\be
\pa_x j^x=-\pa_t j^t=-\pa_x \left(D \N_x j^t \right) \quad \leadsto \quad \pa_t j^t=\pa_x \left(D \N_x j^t \right).
\la{Fick2}
\ee

\ssk
The Ohm law 
\be
j^x=\sqrt{-\bg}\sqrt{-\bg^{tt}\bg^{rr}}\,\bg^{xx}\,F_{tx}\,\vert_{r_\e}=\s^{xx}E_x\,\vert_{r_\e}
\la{ohm}
\ee
contains the expression for the DC conductivity on the horizon:
\be
\s^{xx}=\left. \fr{\sqrt{-\bg}}{\bg_{xx} \sqrt{-\bg_{tt}\,\bg_{rr}}}\,\right|_{r_+}.
\la{DCs}
\ee
From the Einstein relation $D=\s_{DC}/\chi$ we can read off the charge susceptibility $\chi$.\footnote{The Einstein relation can be deduced from the standard arguments \cite{Kovtun:2008kx}. The charge density at the AdS boundary $r\ra \infty$ can be computed from \rf{j0} and the Maxwell field configuration $\d A_t(r)=(\m-Q/r)$ which satisfies the electrostatic Poisson equation 
with the boundary conditions $\d A_t(\infty)=\m$ and $\d A_t(r_+)=0$. Then, at the leading order of $\r(T,\mu)$ and $\m(T)$, $\r=\mathrm{e}^{\Phi(x,y)} Q=\chi \mu$, where $\mu$ is the chemical potential and $\chi$ is the charge susceptibility. Clearly, $\chi=r_+ \mathrm{e}^{\Phi(x,y)}$, $\s_{DC}\equiv \s^{xx}=1$, hence $D=\s_{DC}/\chi$ holds.} Now it is clear that the previously found universal relation $D= d/4\pi T(d-2)$ \cite{Kovtun:2008kx} receives corrections even in the case of the pure Einstein gravity setup. Indeed, in the considered case
\be
D=\fr1{r_+}\,\mathrm{e}^{-\Phi(x,y)}
, \qquad r_+=\fr{4\pi}3 \,T \qquad (l=1),
\la{Dxy}
\ee
and the bound $D\ge d/4\pi T(d-2)$ suggested in \cite{Kovtun:2008kx} (KR bound) is exponentially suppressed. The violation of this bound strongly depends on the choice of inhomogeneity distribution function $\Phi(x,y)$.\footnote{See Appendix B for the details. The distribution functions \rf{solspl1} and \rf{solspl2} correspond to hyperbolic-type surfaces with saddle point(s). Therefore,
we should talk about local domains of violation of the KR universal bound, where the metric potential is positively defined. Note that here we use the notion of distribution function w.r.t. $\Phi(x,y)$ to some extent; the true inhomogeneity distribution density is defined by $\exp(\Phi(x,y))$.}
If the distribution function is positively valued in the selected domain, the damping strength is determined by the local inhomogeneity degree on the horizon surface.

\ssk
\subsection{AdS/CFT calculations of DC conductivity and charge diffusion on the horizon
}

\subsubsection{Derivation with an effective perturbation ansatz}

Now let us compute the transport coefficient in the hydrodynamic limit of AdS/CFT correspondence. Here we follow \cite{Son:2002sd, Policastro:2002se, Maldacena:1998,Maldacena:1997re,Witten:1998qj,Gubser:1998bc}.

\ssk
Following the AdS/CFT prescriptions in computing the retarded Green's function, let us consider a small perturbation of the Maxwell field $\d A_m \ll 1$ over the RN background
\[
\mathrm{d}s^2_0=-f(r)\mathrm{d}t^2+\fr{\mathrm{d}r^2}{f(r)}+r^2\mathrm{e}^{\Phi(x,y)}(\mathrm{d}x^2+\mathrm{d}y^2),\qquad f(r)=\fr{r^2}{l^2}-\fr{\w_2 M}{r}+\fr{k^2 Q^2_e}{2r^2},
\]
\be
A^\0_t(r)=\m-\fr{Q_e}{r}.
\la{RNback}
\ee

\ssk
We choose an effective perturbation ansatz\footnote{Obviously, we lose the symmetry in $x,y$ directions with a general choice of the metric potential $\Phi(x,y)$. So the correct ansatz for the Maxwell field perturbations \rf{Apert} should also depend on $y$ coordinate. However, we handle \rf{Axdef} as an effective perturbation, referring the reader to the next subsection, where we explain the relation between the general perturbation ansatz and what we call here the effective perturbation.}:
\be
\d A_m=(0,0,{A}_x(t,r,x),0).
\la{Apert}
\ee
Then, in the leading order of perturbations, the vector field equation of motion \rf{Meom} turns into
\[
\mathrm{e}^{-\Phi(x,y)}\fr1{r^2 f(r)}\,\pa_t \pa_x A_x+{\mathcal{O}}(A_x^2)=0,\qquad   \mathrm{e}^{-\Phi(x,y)}\fr{f(r)}{r^2}\,\pa_r \pa_x A_x+{\mathcal{O}}(A_x^2)=0,
\]
\be
\mathrm{e}^{-\Phi(x,y)}\fr1{r^2}\left[\pa_r\left(f(r) \pa_r A_x\right)-\fr1{f(r)}\,\pa^2_t A_x \right]+{\mathcal{O}}(A_x^2)=0.
\la{Aperteom}
\ee

\ssk
Setting the effective perturbed field to the plane wave in the $x$ direction
\be
A_x(t,r,x)={\mathfrak{A}}_x (r) \mathrm{e}^{-i\w t+iqx}
\la{Axdef}
\ee
and plugging \rf{Axdef} back to \rf{Aperteom} we conclude that the dynamical equations of the perturbed vector mode in the linear order approximation are compatible in the $q\ra 0$ limit. The dynamics of ${\mathfrak{A}}_x (r)$ is determined by the last equation of \rf{Aperteom}:
\be
\pa_r \left(\bg^{rr}\,\pa_r {\mathfrak{A}}_x \right)-\fr{\w^2}{\bg_{tt}}\, {\mathfrak{A}}_x=0,
\la{frAeq}
\ee
which we rewrite to
\be
{\mathfrak{A}}_x''-\fr{\bg_{rr}}{\bg_{tt}}\,\w^2\, {\mathfrak{A}}_x+\fr{\left(\bg^{rr} \right)'}{\bg^{rr}}\,{\mathfrak{A}}_x'=0.
\la{frAeq2}
\ee
The solution to the latter equation in the $\w \ra 0$ limit is apparent; it is\footnote{One may check 
\[
{\mathfrak{A}}_x(r)=\exp \left(i\a\w \int_r \,dr' \,\bg_{rr}(r') \right),\qquad \a=\fr1{\sqrt{-\bg_{tt}(r_+) \bg_{rr}(r_+)}}
\]
is the solution to \rf{frAeq} satisfying the in-falling boundary condition in the near-horizon limit and the second boundary condition ${\mathfrak{A}}_x \ra 1$ at $r\ra \infty$.}
\be
{\mathfrak{A}}_x(r)=1+\a\, i \w \int_{r}^{\infty}\,\mathrm{d}r'\, \bg_{rr}(r')+{\mathcal{O}}(\w^2).
\la{frAsol}
\ee
with some constant $\a$. This constant is fixed by the requirement of having the in-falling boundary condition near the horizon (see, e.g., Appendix A of \cite{Iqbal:2008by}):
\be
{\mathfrak{A}}_x(r) \propto \exp \left(-i \fr{\w}{4\pi T}\,\ln(r-r_+) \right)=1-i \fr{\w}{4\pi T}\,\ln(r-r_+)+{\mathcal{O}}(\w^2).
\la{Axbc}
\ee
Comparing \rf{frAsol} to \rf{Axbc} in the near-horizon limit we get
\be
\a=\left. \fr1{\sqrt{-\bg_{tt} \bg_{rr}}}\right|_{r_\e}.
\la{alpha}
\ee

\ssk
According to the AdS/CFT recipe \cite{Maldacena:1997re,Maldacena:1998, Gubser:1998bc, Witten:1998qj} we have to substitute the solution to the dynamical bulk equation into the boundary term
\be
S_{\text{on-shell}}=-\fr12 \int_{r \ra \infty} \mathrm{d}^3x\, \sqrt{-\bg}\, A_x \,\bg^{rr} \,\bg^{xx}\, F_{rx}=-\fr12 \int_{r\ra\infty} \mathrm{d}^3 x \,A_x \left(i\a\w \sqrt{-\bg}\,\bg^{xx} \right)A_x,
\la{Sos}
\ee
and to extract the retarded Green's function:
\be
G^R(\w,q \ra 0)=i\a\w \sqrt{-\bg}\,\bg^{xx}.
\la{GR}
\ee
Applying the Kubo formula of the linear response theory leads to the following expression of the DC conductivity near the horizon:
\be
\s^{xx}=\left. \lim_{\w \ra 0} \fr1{\w}\,\text{Im}\, G^R(\w,0)\right|_{r_+}=\left. \fr{\sqrt{-\bg}}{\bg_{xx}\sqrt{-\bg_{tt}\,\bg_{rr}}}\right|_{r_+},
\la{sKubo}
\ee
which coincides with Eq. \rf{DCs} obtained within the stretched horizon approach.

\ssk
The same arguments as before (see footnote 7 to this end) lead to the charge diffusion on the horizon
with the coefficient
\be
D=\fr{3}{4\pi T_{\mathrm{RN}}}\,\mathrm{e}^{-\Phi(x,y)}\quad (l=1),
\la{DxyRN}
\ee
so we arrive at the same conclusions on violation of the KR \cite{Kovtun:2008kx} universal bound as in the previous subsection.

\ssk
\subsubsection{``R-charge'' diffusion pole}

To verify \rf{Dxy} and \rf{DxyRN} we will directly obtain the charge diffusion coefficient adapting the 
computational scheme of \cite{Kovtun:2003wp, Policastro:2002se} to the considered case. Because we are aimed at calculation of the diffusion coefficient from the pole of  Green's function in the momentum representation both temporal and spatial components of a four momentum have to be non-trivial.

\ssk
Following \cite{Kovtun:2003wp, Policastro:2002se} we consider the Maxwell equation
\be
\pa_m (\sqrt{-g}\, g^{kl}\, g^{mn}\,F_{nl})=0
\la{MeomG}
\ee
on a background defined by
\be
\mathrm{d}s^2=g_{tt}(r) \mathrm{d}t^2+g_{rr}(r) \mathrm{d}r^2+\left(g_{xx}(x,y) \mathrm{d}x^2+g_{yy}(x,y) \mathrm{d}y^2 \right).
\la{gendsG}
\ee
We will suppose that $g_{xx}=g_{yy}=r^2 f(x,y)$ and $\sqrt{-g} g^{xx}$, $\sqrt{-g} g^{yy}$ are constants.
After fixing the gauge $A_r=0$ and requiring the most general form of the Maxwell field 
\be
A_m(t,r,x,y)=\int \, \fr{\mathrm{d}\w \mathrm{d}^2q}{(2\pi)^3} \, \mathrm{e}^{-i\w t+iq_{\sx}x+iq_{\sy}y}\,A_m(\w,q_\sx,q_\sy,r),
\la{AansG}
\ee
equation \rf{MeomG} splits into
\be
g^{tt} \w A'_t-g^{xx} q_\sx A'_x-g^{yy} q_\sy A'_y=0,
\la{MeomrG1}
\ee
\be
\pa_r (\sqrt{-g}\, g^{tt} g^{rr} A'_t)-\sqrt{-g}\,g^{tt} g^{xx}(\w q_\sx A_x+q_\sx^2 A_t)-\sqrt{-g}\,g^{tt} g^{yy}(\w q_\sy A_y+q_\sy^2 A_t)=0 ,
\la{MeomtG1}
\ee
\be
\pa_r(\sqrt{-g}\,g^{xx} g^{rr} A'_x)-\sqrt{-g}\,g^{xx} g^{tt}(\w q_\sx A_t+\w^2 A_x)-\sqrt{-g}\,g^{xx} \left[ i\pa_y g^{yy}-g^{yy} q_\sy \right](q_\sx A_y-q_\sy A_x)=0,
\la{MeomxG1}
\ee
\be
\pa_r(\sqrt{-g}\,g^{yy} g^{rr} A'_y)-\sqrt{-g}\,g^{yy} g^{tt}(\w q_\sy A_t+\w^2 A_y)-\sqrt{-g}\,g^{yy} \left[i\, \pa_x g^{xx}-g^{xx} q_\sx \right](q_\sy A_x-q_\sx A_y)=0.
\la{MeomyG1}
\ee
Here the prime denotes the differentiation over $r$.

\ssk

The system of Eqs. \rf{MeomrG1}--\rf{MeomyG1} does not in general admit a decoupled vector field mode. However, we can further specify the ansatz for the Maxwell field to set $q_\sy A_x=q_\sx A_y$. Then Eqs. \rf{MeomrG1}--\rf{MeomyG1} turn into
\be
g^{tt} \w A'_t-{\mathfrak{G}}^{xx,yy}\, q_\sx A'_x=0,
\la{MeomrG1x=y}
\ee
\be
\pa_r (\sqrt{-g}\, g^{tt} g^{rr} A'_t)-\sqrt{-g}\,g^{tt}\, {\mathfrak{G}}^{xx,yy}(\w q_\sx A_x+q_\sx^2 A_t)=0 ,
\la{MeomtG1x=y}
\ee
\be
\pa_r\left(\sqrt{-g}\,{\mathfrak{G}}^{xx,yy} g^{rr} A'_x \right)-\sqrt{-g}\, g^{tt}\,{\mathfrak{G}}^{xx,yy}(\w q_\sx A_t+\w^2 A_x)=0,
\la{MeomxG1x=y}
\ee
where we have introduced the effective metric in the $x,y$ directions
\be
{\mathfrak{G}}^{xx,yy}=g^{xx}+\left(\fr{q_\sy}{q_\sx} \right)^2 g^{yy}.
\la{Gmfrdef}
\ee
The structure of \rf{MeomrG1x=y}--\rf{MeomxG1x=y} is similar to the corresponding system of equations of \cite{Kovtun:2003wp} (or \cite{Policastro:2002se}), the solution to which leads to the expression for the charge diffusion coefficient.

\ssk
Following \cite{Policastro:2002se, Kovtun:2003wp}, let us join \rf{MeomrG1x=y}, \rf{MeomtG1x=y} into a single equation,
\be
\fr{\mathrm{d}}{\mathrm{d}r} \left[\fr{\pa_r (\sqrt{-g}\, g^{tt} g^{rr} A'_t)}{\sqrt{-g}\,g^{tt} {\mathfrak{G}}^{xx,yy}} \right]-\left(\fr{g^{tt}}{{\mathfrak{G}}^{xx,yy}}\, \w^2+q_\sx^2 \right) A'_t=0.
\la{MeomtG2x=y}
\ee
On account of \rf{MeomrG1x=y}, Eqs. \rf{MeomtG1x=y} and \rf{MeomxG1x=y} become equivalent, so it is enough to solve \rf{MeomtG2x=y} for $A'_t$. 

\ssk
Now we specify the background metric (see \rf{dspliso}):
\be
g_{tt}=-\left(\fr{r^2}{l^2}-\fr{\w_2 M}{r} \right)\equiv -f(r),\qquad g_{tt}g_{rr}=-1,\quad {\mathfrak{G}}^{xx,yy}=\fr{\mathrm{e}^{-\Phi(x,y)}}{r^{2}}\left(1+\left(\fr{q_\sy}{q_\sx} \right)^2 \right)\equiv \fr{\g}{r^2} \mathrm{e}^{-\Phi(x,y)}.
\la{specGx=y}
\ee
Searching for solutions to the second order w.r.t. $A'_t$ Eq. \rf{MeomtG2x=y} is simplified with introducing new radial variable $u=r_+/r$. Setting $l=r_+=1$ and $\w_2 M/r_+=1$, the gravitational background becomes
\be
\mathrm{d}s^2=\fr1{u^2}\left[-f(u)\mathrm{d}t^2+\fr{\mathrm{d}u^2}{f(u)}+\mathrm{e}^{\Phi(x,y)}(\mathrm{d}x^2+\mathrm{d}y^2)\right],\qquad f(u)=1-u^3 .
\la{dsu}
\ee
The metric potential $\Phi(x,y)$ still obeys the elliptic wave equation \rf{rhoeqpl}; the set of expressions \rf{specGx=y} turns into
\be
g_{tt}=-(1-u^3)/u^2,\quad g_{tt}g_{uu}=-1/u^4,\quad {\mathfrak{G}}^{xx,yy}=\g\, u^2 \mathrm{e}^{-\Phi(x,y)},
\la{specGx=yu}
\ee
and $A_t(r) \ra A_t(u)$. Then \rf{MeomtG2x=y} becomes
\be
\fr{\mathrm{d}}{\mathrm{d}u}\left(f(u) \fr{\mathrm{d}}{\mathrm{d}u}A'_t \right)+\left(\fr{\w^2}{f(u)}-\mathrm{e}^{-\Phi}\, \tilde{q}^2 \right) A'_t=0,
\la{MeomtG2x=yu}
\ee
where the prime now corresponds to the derivative over $u$. It is also convenient to turn to the rescaled momentum $\tilde{q}^2=\g q^2_\sx=q^2_\sx+q^2_\sy$ and to introduce the dimensionless energy and momentum,
\be
{\mathfrak{w}}=\fr{3\w}{4\pi T}, \qquad {\mathfrak{q}}=\fr{3 \tilde{q}}{4\pi T}.
\la{wqdimles}
\ee

\ssk
Equation \rf{MeomtG2x=yu} is perturbatively solved in the small ${\mathfrak{w}}$, ${\mathfrak{q}}^2$ approximation \cite{Policastro:2002se, Kovtun:2003wp} with the following ansatz:
\be
A'_t=C\,(1-u)^{-i{\mathfrak{w}}/3} \,(1+{\mathfrak{w}} F_1(u)+{\mathfrak{q}}^2 G_1(u,x,y))+{\mathcal{O}}({\mathfrak{w}}^2,{\mathfrak{w}}{\mathfrak{q}}^2,{\mathfrak{q}}^3), 
\la{A'tans}
\ee
satisfying the near-horizon incoming wave boundary condition. The functions $F_1(u)$, $G_1(u,x,y)$ entering \rf{A'tans} are required to be regular at the horizon $u=1$. 

\ssk
Plugging \rf{A'tans} back to \rf{MeomtG2x=yu} one gets 
\be
F_1(u)=\fr{i}{\sqrt{3}}\left[ \tan^{-1}\left(\fr{1+2u}{\sqrt{3}} \right)+\fr1{2 \sqrt{3}}\,\ln(1+u+u^2) \right],
\la{F1u}
\ee
\be
G_1(u,x,y)=-\fr{2}{\sqrt{3}}\,\mathrm{e}^{-\Phi} \tan^{-1}\left(\fr{1+2u}{\sqrt{3}} \right).
\la{G1u}
\ee

\ssk
The overall constant $C$ of \rf{A'tans} is fixed from Eq. \rf{MeomtG1x=y} written in $u$ variable
\be
A''_t=\left(\fr{4\pi T}{3} \right)^2 \fr{\mathrm{e}^{-\Phi}}{f(u)}\left({\mathfrak{w}}{\mathfrak{q}} A_x+{\mathfrak{q}}^2 A_t \right).
\la{MeomtG1u}
\ee
Taking the limit $u \ra 0$ we get
\be
C=\left. \fr{1}{i{\mathfrak{w}}-\mathrm{e}^{-{\Phi}} {\mathfrak{q}}^2} \left(\fr{4\pi T}{3} \right)^2 \left\{ \mathrm{e}^{-\Phi} \left({\mathfrak{w}}{\mathfrak{q}} A_x+{\mathfrak{q}}^2 A_t \right) \right\} \right|_{u=0},
\la{Csol}
\ee
and the pole in $C$ is the diffusion pole with the diffusion ``constant''
\be
D=\fr{3}{4\pi T}\, \mathrm{e}^{-\Phi(x,y)}.
\la{Du}
\ee

\ssk
In the case of a perturbation of the Maxwell mode over the RN background
\be
\mathrm{d}s^2=\fr1{u^2}\left[-f(u)\mathrm{d}t^2+\fr{\mathrm{d}u^2}{f(u)}+\mathrm{e}^{\Phi(x,y)}(\mathrm{d}x^2+\mathrm{d}y^2) \right],\quad f(u)=1-u^3-\fr{k^2 Q^2}2\,u^3(1-u),
\la{RNLu}
\ee
\be
A^\0_t=\m-qu\,
\la{AtLu}
\ee
the solution to \rf{MeomtG2x=yu} may be found \cite{Son:2006em} from 
\be
A'_t=C\,f(u)^{-i{\mathfrak{w}}/3} \,(1+{\mathfrak{w}} \tilde{F}_1(u)+{\mathfrak{q}}^2 \tilde{G}_1(u,x,y))+{\mathcal{O}}({\mathfrak{w}}^2,{\mathfrak{w}}{\mathfrak{q}}^2,{\mathfrak{q}}^3), 
\la{A'tansRN}
\ee
that after the same steps as before leads to the diffusion pole at $i{\mathfrak{w}}=\mathrm{e}^{-\Phi}{\mathfrak{q}}^2/(1-k^2Q^2/6)$. On account of the relation between the Hawking temperatures of the RN and neutral BHs
\be
T_{\mathrm{RN}}=T\left(1-\fr{k^2 Q^2}6 \right),
\la{TRNT}
\ee
we end up with the diffusion coefficient 
\be
D=\fr{3}{4\pi T_{\mathrm{RN}}}\,\mathrm{e}^{-\Phi(x,y)}.
\la{DuRN}
\ee

\ssk
Comparing the values of the charge diffusion coefficients \rf{Du}, \rf{DuRN} to that of previously obtained from the membrane paradigm (Eq. \rf{Dxy}) and from the effective Maxwell field perturbations ansatz (Eq. \rf{DxyRN}) we note that the results coincide. 
A formal correspondence between variables in the former and in the latter computing schemes is as follows:
\be
g^{xx} \Lra {\mathfrak{G}}^{xx,yy},\qquad \Phi(x,y) \Lra \Phi(x,y),\qquad q \Lra \tilde{q},\qquad T \Lra T_{\mathrm{RN}}.
\la{gqcorr}
\ee

\ssk
Note also that the dispersion relation, following from the second Fick law \rf{Fick2},
\be
i\w\sim D(q^2_i+iq_i \pa_i \Phi),\qquad i=x,y
\la{Disper}
\ee
contains the term linear in the momentum with the gradient of the metric potential $\Phi(x,y)$. This contribution to the diffusion pole has to appear after relaxing the imposed condition $q_\sy A_x=q_\sx A_y$. Such a generalisation for the Maxwell field perturbation is beyond the scope of the paper and will be done elsewhere. Below we will see the appearance of the linear over momentum term in the diffusion pole of gravitational perturbations over AdS$_4$ BH background.

\section{Shear viscosity and $\eta/s$ ratio}

In this part of the paper we compute the $\eta/s$ ratio for the planar BH solution with inhomogeneous 
horizon surface. As in the previously considered case of charge diffusion, calculations in this section are divided into two parts.  We perform quick effective computations first, after that we will turn to a more rigorous computational scheme to verify the result.  

\subsection{A quick derivation}

Let us perturb the background \rf{dspliso} with $h_{xy} \ll 1$ mode \cite{Policastro:2002se}, i.e.
\be
\mathrm{d}s^2=-f(r)\mathrm{d}t^2+\fr{\mathrm{d}r^2}{f(r)}+r^2 \mathrm{e}^{\Phi(x,y)}(\mathrm{d}x^2+\mathrm{d}y^2)+2r^2 h_{xy}(t,r,x)\mathrm{d}x \mathrm{d}y,\quad f(r)=\left(\fr{r^2}{l^2}-\fr{\w_2 M}{r} \right).
\la{dsh}
\ee
Expanding the Einstein equation $R_{mn}=\L g_{mn}$ to the linear order in the fluctuations we get
\[
R_{tt}(\bg+\d g)=\L \bg_{tt}+{\mathcal{O}}(h^2),\quad R_{rr}(\bg+\d g)=\L \bg_{rr}+{\mathcal{O}}(h^2), 
\]
\[
R_{xx}(\bg+\d g)=\L \bg_{xx}-\fr1{2\L} \pa_y \left(\mathrm{e}^{-\Phi}\,\pa_x h_{xy} \right)+{\mathcal{O}}(h^2),\quad R_{yy}(\bg+\d g)=\L\bg_{yy}-\fr1{2\L} \pa_y \left(\mathrm{e}^{-\Phi}\,\pa_x h_{xy} \right)+{\mathcal{O}}(h^2),
\]
\[
R_{ty}(\bg+\d g)=-\fr1{2\L} \pa_t \left(\mathrm{e}^{-\Phi}\,\pa_x h_{xy} \right)+{\mathcal{O}}(h^2),
\]
\[
R_{ry}(\bg+\d g)=-\fr1{2\L} \pa_r \left(\mathrm{e}^{-\Phi}\,\pa_x h_{xy} \right)+{\mathcal{O}}(h^2),
\]
\[
R_{xy}(\bg+\d g)-\L(\bg+\d g)_{xy}=\left(\L r^2+f +r\pa_r f \right)h_{xy}+\fr12\pa_r \left(r^2 f \,\pa_r h_{xy} \right)- \fr{r^2}{2f}\,\pa^2_t h_{xy}+{\mathcal{O}}(h^2).
\]
where we have used Eq. \rf{rhoeqpl}. Taking
\be
h_{xy}(t,r,x)={\mathfrak{h}}_{xy}(r)\mathrm{e}^{-i\w t+iqx}
\la{hxyr}
\ee
one may notice that the resulted system of equations is compatible to the zero momentum limit $q\ra 0$, which is suitable for our aims.
Therefore, to compute the shear viscosity we have to solve the equation for ${\mathfrak{h}}_{xy}(r)$ first.

\ssk
Plugging \rf{hxyr} into the $xy$ part of the perturbed Einstein equation one gets
\be
\fr1{r^2 \mathrm{e}^\Phi}\pa_r\left(r^2 \mathrm{e}^\Phi \,f\,\pa_r {\mathfrak{h}}_{xy} \right)+\fr{\w^2}{f}\,{\mathfrak{h}}_{xy}+\left(\fr{2\pa_r f}{r}+\fr{2f}{r^2}-6 \right) {\mathfrak{h}}_{xy}=0,
\la{hxyreq}
\ee
which on account of the explicit expression for $f(r)$ can be written down in the following form:
\be
\fr{1}{\sqrt{-\bg}}\pa_r \left(\sqrt{-\bg}\, \bg^{rr} \pa_r {\mathfrak{h}}_{xy} \right)+\fr{1}{\sqrt{-\bg}}\pa_t \left(\sqrt{-\bg}\, \bg^{tt} \pa_t h_{xy} \right)=0.
\la{hxyrcov}
\ee
Then the resulting equation for ${\mathfrak{h}}_{xy}$ 
\be
\pa^2_r {\mathfrak{h}}_{xy}-\fr{\bg_{rr}}{\bg_{tt}}\,\w^2 \,{\mathfrak{h}}_{xy}+\fr1{\bg^{rr} \sqrt{-\bg}}\pa_r \left(\bg^{rr} \sqrt{-\bg} \right)\,\pa_r {\mathfrak{h}}_{xy}=0
\la{hxyrcov1}
\ee
coincides with that obtained in \cite{Natsuume:2010ky} (cf. Eq. (2.19) therein).

\ssk
Solution to Eq. \rf{hxyrcov1} satisfying the boundary condition ${\mathfrak{h}}_{xy} \ra 1$, $r \ra \infty$ is
\be
{\mathfrak{h}}_{xy}(r)=\exp \left(i\a\w \int_r \,\mathrm{d}r'\,\fr{\bg_{rr}(r')}{\sqrt{-\bg (r')}} \right). 
\la{hxysol}
\ee
Again, the constant $\a$ is fixed by the in-falling boundary condition at the horizon
\be
h_{xy}(r) \propto \exp \left(-i \fr{\w}{4\pi T} \ln(r-r_+) \right)=1-i\fr{\w}{4\pi T} \ln(r-r_+)+{\mathcal{O}}(\w^2).
\la{hxybc}
\ee
Comparing \rf{hxysol} to \rf{hxybc} in the near-horizon limit results in
\be
\a=-4G\,s
\la{alphah}
\ee
with the entropy density
\be
s=\fr{\mathrm{d}A}{4G}=\left. \fr{\left(\bg_{xx} \bg_{yy} \right)^{1/2}}{4G}\right|_{r_+}=s_{0}\mathrm{e}^{\Phi}.
\la{sdef}
\ee
Here we have denoted the entropy of a BH with isotropic homogeneous horizon $s_0$. Clearly, the entropy density $s$ becomes a function of the local distribution $\Phi(x,y)$.

\ssk
The on-shell action for the perturbed gravity mode is as follows:
\be
S_{\text{on-shell}}=\fr{1}{16 \pi G} \,\int_{r \ra \infty} \mathrm{d}x^3\,\fr12 \, \sqrt{-\bg}\,\bg^{rr}\,{\mathfrak{h}}_{xy} \pa_r {\mathfrak{h}}_{xy}=\fr{1}{16 \pi G} \,\int_{r \ra \infty} \mathrm{d}x^3\,\fr12 {\mathfrak{h}}_{xy} \left(i\a\w \right) {\mathfrak{h}}_{xy} .
\la{Shos}
\ee
Plugging the retarded Green function 
\be
G^R(\w,q \ra 0)=i\,\fr{\w s}{4\pi}
\la{GRh}
\ee
into the Kubo formula leads to the well-known expression \cite{Policastro:2002se}
\be
\eta/s=\fr{1}{4\pi}.
\la{shear0}
\ee
Now, once we recover the KSS $\eta/s_0$ ratio \cite{Kovtun:2003wp}, from \rf{sdef} we get 
\be
\eta/s_0=\fr1{4\pi}\mathrm{e}^{-\Phi(x,y)}.
\la{shear}
\ee

\ssk
Therefore, as in the case of charge diffusion near the BH inhomogeneous horizon, the KSS universal bound relation $\eta/s_0 \ge 1/4\pi$ receives exponential suppression. The range of its violation and the possibility to violate this bound at all strongly depend on the local properties of the metric potential $\Phi(x,y)$ (see the discussion around footnote 8). Equation \rf{shear} also follows from computations of the shear viscosity in the RN background metric \rf{RNback} or its magnetically charged cousins (solutions \rf{RNsolem}--\rf{RNAsolem} with $K=0$).

\subsection{Diffusion pole in 4D AdS BH with Liouville mode}

Now we will check the relation \rf{shear} by computing the shear viscosity from the diffusion pole of gravitational perturbation Green's function \cite{Kovtun:2003wp, Policastro:2002se}. It is convenient to turn to the AdS$_4$ BH background in $u$ radial variable to this end:
\be
\mathrm{d}s^2=\fr1{u^2}\left[-f(u)\mathrm{d}t^2+\fr{\mathrm{d}u^2}{f(u)}+\mathrm{e}^{\Phi(x,y)}(\mathrm{d}x^2+\mathrm{d}y^2)\right],\qquad f(u)=1-u^3.
\la{AdS4einLu}
\ee
Equation \rf{AdS4einLu} defines the background metric $\bar{g}_{mn}$, and we choose the perturbation of the gravitational field $g_{mn}=\bar{g}_{mn}+h_{mn}$, $h_{mn} \ll 1$ in the following form:
\be
h_{mn}=\{h_{ty}(t,u,x,y)/u^2,h_{xy}(t,u,x,y)/u^2\},
\la{hansu}
\ee
with the other components of $h_{mn}$ equal to zero (we work in the $h_{mu}=0$ gauge).

\bsk
Imposing the elliptic wave equation on the metric potential $\Phi(x,y)$, we get the following equations for the non-trivial components of $h_{mn}$ up to the second order in perturbations:
\be
\mathrm{e}^{-\Phi}\pa_t \pa_y h_{ty}=0,\qquad  \pa_x \Phi \,\pa_y h_{xy}+\pa_y \Phi \,\pa_x h_{xy}-2\pa_x\pa_y h_{xy}+\fr{\mathrm{e}^{\Phi}}{f(u)}\,\pa_y \Phi \, \pa_t h_{ty}=0,
\la{ttxxL}
\ee
\be
\pa_x \Phi \,\pa_y h_{xy}+\pa_y \Phi \,\pa_x h_{xy}-2\pa_x\pa_y h_{xy}+\fr{\mathrm{e}^{\Phi}}{f(u)}\,(\pa_y \Phi \, \pa_t h_{ty}-2\pa_t\pa_y h_{ty})=0,
\la{yyL}
\ee
\be
\mathrm{e}^{-\Phi}\left(\pa_u \pa_y h_{ty}-\fr{f'(u)}{f(u)}\,\pa_y h_{ty} \right)=0,\qquad \pa_x\pa_y h_{ty}+\pa_t \pa_y h_{xy}-\pa_y \Phi \,\pa_x h_{ty}=0\,\qquad  \mathrm{e}^{-\Phi}\pa_u\pa_y h_{xy}=0,
\la{txtyuxL}
\ee
\be
\pa_u \pa_x h_{xy}-\fr{\mathrm{e}^{\Phi}}{f(u)}\,\pa_t\pa_u h_{ty}=0,
\la{uyL}
\ee
\be
\pa_u^2 h_{ty}-\fr{2}{u}\,\pa_u h_{ty}+\fr{\mathrm{e}^{-\Phi}}{f(u)}\left(\pa^2_x h_{ty}-\pa_x \Phi\,\pa_x h_{ty}-\pa_t\pa_x h_{xy}\right)=0,
\la{tyL}
\ee
\be
\pa_u^2 h_{xy}-\fr{3-f(u)}{u f(u)}\,\pa_u h_{xy}+\fr1{f^2(u)}\left(\pa_t\pa_x h_{ty}-\pa_x \Phi\,\pa_x h_{ty}-\pa_t^2 h_{xy}\right)=0.
\la{xyL}
\ee

\ssk
Equations \rf{ttxxL}--\rf{txtyuxL} are consistent once $\pa_y h_{ty}=0$, $\pa_y h_{xy}=0$ and $\pa_y \Phi=0$. The consistency conditions select ``chiral'' in $x,y$ plane modes with $h_{mn}=\{h_{ty}(t,u,x)/u^2,h_{xy}(t,u,x)/u^2 \}$ and $\Phi(x,y) \ra\Phi(x)$. However, the ``chirality'' condition on the metric potential is stringent enough: together with the elliptic wave equation it narrows the choice of non-trivial metric potential to $\Phi(x)=\a x+\b$. Hence, we have to relax $\pa_y \Phi(x,y)=0$.

\ssk
To relax the ``chirality'' condition on the metric potential $\Phi(x,y)$ we add other ``chiral'' perturbation modes in $y$ direction, $H_{mn}=\{ H_{tx}(t,u,y)/u^2,H_{xy}(t,u,y)/u^2\}$, $H_{mn} \ll 1$. As a result we have the following system of non-trivial equations for the perturbed metric $g_{mn}=\bar{g}_{mn}+h_{mn}+H_{mn}$ in the first order in the perturbations:
\be
\pa_x h_{xy}'-\fr{\mathrm{e}^{\Phi}}{f(u)}\,\pa_t h'_{ty}=0,\qquad \pa_y H_{xy}'-\fr{\mathrm{e}^{\Phi}}{f(u)}\,\pa_t H'_{tx}=0,
\la{uyuxL2}
\ee
\be
H''_{tx}-\fr2{u}\,H'_{tx}+\fr{\mathrm{e}^{-\Phi}}{f(u)} \left(\pa^2_y H_{tx}-\pa_y \Phi (\pa_y H_{tx}-\pa_x h_{ty})-\pa_t \pa_y H_{xy} \right)=0,
\la{txL2}
\ee
\be
h''_{ty}-\fr2{u}\,h'_{ty}+\fr{\mathrm{e}^{-\Phi}}{f(u)} \left(\pa^2_x h_{ty}-\pa_x \Phi (\pa_x h_{ty}-\pa_y H_{tx})-\pa_t \pa_x h_{xy} \right)=0,
\la{tyL2}
\ee
\[
h''_{xy}-\fr{3-f(u)}{u f(u)}h'_{xy}+\fr1{f^2(u)}\left(\pa_t\pa_x h_{ty}-\pa_x\Phi\,\pa_t h_{ty}-\pa^2_t h_{xy}\right)
\]
\be
+H''_{xy}-\fr{3-f(u)}{u f(u)}H'_{xy}+\fr1{f^2(u)}\left(\pa_t\pa_y H_{tx}-\pa_y\Phi\,\pa_t H_{tx}-\pa^2_t H_{xy}\right)=0,
\la{xyL2}
\ee
while the $tt,tu,uu$ components of the AdS$_4$ Einstein equation turn to identities.
Additionally we have the following consequences of the $xx$ and $yy$ components of the AdS$_4$ Einstein equation:
\be
\pa_y \Phi \,\pa_x h_{xy}+\pa_x \Phi \,\pa_y H_{xy}=0,\qquad \pa_y \Phi \,\pa_t h_{ty}-\pa_x\Phi \,\pa_t H_{tx}=0.
\la{xx+yy,xx-yy}
\ee

\ssk
Turning to the Fourier modes 
\[
h_{mn}(t,u,x)=\int\,\fr{\mathrm{d}\w_\sx \mathrm{d}q_\sx}{(2\pi)^2}\,\mathrm{e}^{-i\w_\sx t+iq_\sx x}\, h_{mn}(\w_\sx,q_\sx,u),
\]
\be
H_{mn}(t,u,y)=\int\,\fr{\mathrm{d}\w_\sy \mathrm{d}q_\sy}{(2\pi)^2}\,\mathrm{e}^{-i\w_\sy t+iq_\sy y}\, H_{mn}(\w_\sy,q_\sy,u),
\la{hHansFx2}
\ee
one may check that the consistency of \rf{xyL2} with \rf{uyuxL2}--\rf{tyL2} requires
\be
\pa_y \Phi\,\fr{\w_\sy}{q_\sy}\, q_\sx h_{ty}+\pa_x\Phi \,\fr{\w_\sx}{q_\sx}\, q_\sy H_{tx}=0.
\la{xyL2c2}
\ee
Equations \rf{xx+yy,xx-yy} in the momentum representation are
\be
\pa_y\Phi\, q_\sx h_{xy}+\pa_x \Phi\, q_\sy H_{xy}=0,\qquad \pa_y\Phi\,\w_\sx h_{ty}-\pa_x \Phi\,\w_\sy H_{tx}=0,
\la{xxyy2}
\ee
so Eqs. \rf{xyL2c2}, \rf{xxyy2} turn to identities (in the linear order in $h_{mn}$, $H_{mn}$) if
\be
h_{mn}=\{h_{ty},h_{xy}\} \ll \pa_y \Phi,\qquad H_{mn}=\{H_{tx},H_{xy}\} \ll \pa_x \Phi.
\la{hHPhi}
\ee
Since the perturbation modes $h_{mn}$, $H_{mn}$ are small, Eqs. \rf{hHPhi} can always be fulfilled for a general choice of the metric potential $\Phi(x,y)$.

\ssk
Once the restrictions \rf{hHPhi} are taken into account eqs. \rf{uyuxL2}--\rf{xyL2} form the consistent set of equations, and solutions to \rf{txL2}, \rf{tyL2} are (on account of \rf{uyuxL2}) solutions to \rf{xyL2}.

\ssk
We will solve Eqs. \rf{uyuxL2}--\rf{tyL2} similarly to \cite{Kovtun:2003wp, Policastro:2002se} (see also \cite{Herzog:2002fn}). From \rf{uyuxL2}--\rf{tyL2} it follows 
\be
\fr{\mathrm{d}}{\mathrm{d}u} \left(f(u)\,(H''_{tx}-\fr2{u}\,H'_{tx})\right)+\left(\fr{\w^2_\sy}{f(u)}-\mathrm{e}^{-\Phi}(q^2_\sy+iq_\sy \pa_y\Phi) \right)H'_{tx}=0,
\la{Htxtosol}
\ee
\be
\fr{\mathrm{d}}{\mathrm{d}u} \left(f(u)\,(h''_{ty}-\fr2{u}\,h'_{ty})\right)+\left(\fr{\w^2_\sx}{f(u)}-\mathrm{e}^{-\Phi}(q^2_\sx+iq_\sx \pa_x\Phi) \right)h'_{ty}=0.
\la{htytosol}
\ee
Let us focus on \rf{Htxtosol}. We will try the following ansatz:
\be
H'_{tx}=C(1-u)^{-i{\mathfrak{w}}/3}\left(F_0(u)+{\mathfrak{w}}_\sy F_1(u)+\mathrm{e}^{-\Phi}\left(i\mathfrak{d}_\sy{\mathfrak{q}}_\sy G_1(u)+{\mathfrak{q}}^2_\sy G_2(u)\right) \right)+{\mathcal{O}}({\mathfrak{w}}^2,{\mathfrak{w}}{\mathfrak{q}}^2,{\mathfrak{q}}^3)
\la{Htxans}
\ee
with functions $F_{0,1}(u)$, $G_{1,2}(u)$ regular at the horizon $u=1$. Here $\mathfrak{d}_\sy=3 \pa_y \Phi/(4\pi T)$ and ${\mathfrak{w}}_\sy=3\w_\sy/(4\pi T)$, ${\mathfrak{q}}_\sy=3 q_\sy/(4\pi T)$. Substituting the ansatz \rf{Htxans} in \rf{Htxtosol} we get
\be
F_0(u)=u^2,\qquad G_1(u)=G_2(u)=\fr13(u^2-u),
\la{HtxsolF0G}
\ee
\be
F_1(u)=i \left(u-u^2+\fr{u^2}{\sqrt{3}} \tan^{-1} \left(\fr{1+2u}{\sqrt{3}} \right)-\fr{u^2}{6}\ln (1+u+u^2) \right).
\la{HtxsolF1}
\ee
The constant $C$ is fixed from the original equation \rf{txL2} at the boundary of AdS space
\be
C=\fr1{\fr13 \mathrm{e}^{-\Phi}({\mathfrak{q}}_\sy^2+i\mathfrak{d}_\sy{\mathfrak{q}}_\sy)-i{\mathfrak{w}}_\sy}\,\mathrm{e}^{-\Phi} \left((q^2_\sy+i\pa_y \Phi \,q_\sy)H_{tx}+\w_\sy q_\sy H_{xy}\right) \Bigg|_{u=0}.
\la{CHtx}
\ee
Therefore, the diffusion coefficient in $y$ direction is equal to
\be
D=\fr1{4\pi T}\,\mathrm{e}^{-\Phi}.
\la{DHtx}
\ee
Since $D=\eta/(\e+P)$ and $\e+P=s_0 T$,\footnote{Thermodynamic equilibrium is defined by $\e_0+P_0=s_0 T_0$ at zero metric potential. Turning the metric potential on one gets
$\mathrm{e}^{-\Phi}(\e_0+P_0)=s_0 \mathrm{e}^{-\Phi}T_0$, that is, $\e+P=s_0 T$.}
\be
\fr{\eta}{s_0}=\fr1{4\pi}\,\mathrm{e}^{-\Phi}.
\la{HtxKSS}
\ee
Equation \rf{htytosol} solves in the same way, so the diffusion coefficient in $x$ directions is also defined by \rf{DHtx}. Therefore, Eq. \rf{shear} holds.

\ssk
We pay the reader's attention on the occurrence of linear in the corresponding momentum term in the diffusion poles of gravitational perturbations along $x,y$ directions. As we have pointed out above, this term results from the diffusion law in inhomogeneous media (see Eq. \rf{Disper}).

\section{Comments and speculations}

\subsection{Comments on transport coefficients for non-planar inhomogeneous horizons}

Calculations performed in previous sections may be extended to the case of black hole solutions with non-zero constant curvature horizons (solutions \rf{RNsolem}--\rf{RNAsolem} with $K=\pm 1$). In this cases one also gets the exponential suppression in formulae for the diffusion coefficient and for the $\eta/s_0$ ratio, Eqs. \rf{Dxy} and \rf{shear}. Formally the results are the same; however, there are differences in compare to the case of a planar inhomogeneous horizon. 

\ssk
Looking at solutions to the elliptic Liouville equation (some of which are borrowed from \cite{Crowdy97} and presented in Appendix B), one can notice that $\exp (\pm\Phi(x,y))$ contains in general different singularities: poles of complex functions entering the solution for the spherical-type horizon and zeros of their derivatives/specific combination (cf. Crowdy's solution \rf{Cr}), or simultaneous zeros of a function derivatives in $x,y$ directions/zeros of functions (cf. Popov's solution \rf{Po} \cite{Popov93}) for the hyperbolic-type horizon. These obstacles should be taken into account upon the choice of trial functions for the Liouville mode: various singularities of functions and their derivatives have to be avoided to keep a well-defined range of physically accepted values of $\g_{ij}$ components of the metric tensor and their inverse.
Note that in both cases (with $K=\pm 1$) $\exp(-\Phi(x,y))$ takes the whole range of values (smaller and greater than one), hence the exponential suppression of $D$ and $\eta/s_0$ with violation of the universal bounds takes place within the local domains of $\exp(-\Phi(x,y))<1$.

\subsection{Comments on the Liouville field in condensed matter physics}

The Liouville equation has been widely recognised in 2D QFT (see, e.g., \cite{abdalla01,LAGG91} for  comprehensive reviews). On the condensed matter physics side the appearance of the Liouville field theory may be found in the description of disordered charged media at the strong coupling limit \cite{Kogan:1996wk} and in the consideration of particle motion in a random potential \cite{CD01} or in a diffusion process of a random walk particle in the $\d$-potential \cite{Ferrari:2006ap, Ferrari:2009vi} (see also \cite{HBA02} for a review of diffusion processes in disordered media). These observations give us evidence to interpret the parameter of inhomogeneity of the horizon $\Phi(x,y)$ as the inhomogeneity degree in the dual strongly coupled effective media related to its disorder and the degree of chaotisation. We hope it opens a new prospect in searching for a holographic description of such CMP models in terms of the gauge/gravity duality.\footnote{See also \cite{Germani:2013sra,Dvali:2012en} as an interplay between the Bose--Einstein condensate, BH physics and the Liouville theory.}

\subsection{Comments on higher-dimensional generalisation of the solutions}

Higher-dimensional generalisation of the solutions \rf{RNsolem}--\rf{RNAsolem} is easy to derive on account of the previously found solution for an electrically charged AdS$_{n+1}$ black hole \cite{Chamblin:1999tk}. Adapting to our case this solution transforms into
\be
\mathrm{d}s^2=-f(r)\mathrm{d}t^2+\fr{\mathrm{d}r^2}{f(r)}+r^2\left(\mathrm{e}^{\Phi(x,y)}(\mathrm{d}x^2+\mathrm{d}y^2)+\sum_{i=1}^{n-3} (\mathrm{d}x^i)^2 \right),
\la{AdSD}
\ee
\be
f(r)=\fr{r^2}{l^2}-\fr{\w_{n-1} M}{r^{D-3}} +\fr{k^2 Q^2}{r^{2n-4}}+K\,\qquad A_t(r)=\m-\sqrt{\fr{n-1}{2(n-2)}}\,\fr{Q}{r^{n-2}},
\la{f[r]D}
\ee
where the metric potential $\Phi(x,y)$ satisfies the elliptic Liouville equation
\[
\fr{\pa^2 \Phi}{\pa x^2}+\fr{\pa^2 \Phi}{\pa y^2}+2K\mathrm{e}^{\Phi(x,y)}=0,\qquad K=0,\pm 1 .
\]

\ssk
Applying the technique has been used in computing transport coefficients, one may found the effect of anisotropy in $n$-dimensional effective dual media. For instance, in 5D case we get two different conductivities: 
\be
\s^{xx}=\s^{yy}=r_+,\qquad \s^{zz}=r_+ \mathrm{e}^{\Phi},
\la{sxyz}
\ee
and two different diffusion coefficients:
\be
D_{x}=D_{y}=\fr1{2 r_+} \mathrm{e}^{-\Phi},\qquad D_{z}=\fr1{2 r_+}=\fr1{2 \pi T},
\la{Dxyz}
\ee
one of which is at the KR \cite{Kovtun:2008kx} bound value; the other one is exponentially suppressed. The ratio $\s^{zz}/\s^{xx}=\mathrm{e}^\Phi$ depends on the degree of inhomogeneity, determined by the metric potential, and it is always smaller than one inside domains of $\exp(\Phi(x,y))<1$, where the KR bound holds.
Similar behaviour of $\s^{zz}/\s^{xx}=1/\mathcal{H}(r)$ was early established in the strongly coupled anisotropic plasma \cite{Erdmenger:2010xm, Mateos:2011ix, Mateos:2011tv, Rebhan:2011vd, Critelli:2014kra} with the anisotropy function $\mathcal{H}(r)$, so we observe the formal correspondence $\exp(-\Phi(x,y)) \rightleftharpoons \mathcal{H}(r)$ between the (inverse of) inhomogeneity distribution function on isotropic horizon and the anisotropy function of a homogeneous horizon surface.\footnote{Note, however, the difference between our model and that of \cite{Rebhan:2011vd}: the choice $\mathcal{H}(r)>1$ violates the KR/KSS bounds (see, e.g., \cite{Rebhan:2011vd, Mamo:2012sy}), while to reach $\s^{zz}/\s^{xx} <1$ in our case the KR bound should be preserved.}

\ssk
\subsection{Fitting to RHIC and LHC data}

The combined analysis of all data of high-ion collisions measured at RHIC and at LHC gives the following experimental restrictions on the $\eta/s$ value \cite{Song:2012ua,Gale:2012rq}:
\be
\eta/s \sim 0.12 \,\,\text{(RHIC)},\qquad \eta/s \sim 0.2 \,\,\text{(LHC)}.
\la{etasexp}
\ee
In the following we will focus on the LHC result. Also we will accept the KSS value for the lower bound, $\eta/s_0 \sim 0.08$. 

\ssk
According to our calculations, the $\eta/s$ ratio, measured in $s_0$ units, is not a constant anymore; it is a function of the Liouville mode $\Phi(x,y)$ (cf. \rf{shear}) with natural ``boundary condition''
\be
\fr{\eta}{s}\equiv\fr{\eta}{s_0}(\Phi)
\,\,\, \xrightarrow[\Phi=0]{}\,\,\, \fr{\eta}{s_0}.
\la{etabc}
\ee
Hence, in our interpretation a wide range of experimentally fixed values of $\eta/s$ \rf{etasexp} is an impact of the local inhomogeneity distribution in quark--gluon plasma. Then the KSS bound value corresponds to the QGP near-equilibrium isotropic homogeneous state.

\ssk
Now our aim is to find a shape of the metric potential $\Phi(x,y)$ which will satisfy: (1) the Liouville equation; and (2) the b.c. \rf{etabc} and the upper value bound $\eta/s \sim 0.2$. From the discussion in Sect. 5.1. we have only a chance to realise the experimentally estimated upper bound value with $\Phi(x,y)$ unbounded from below and having the upper bound to be equal to zero. The simplest way to realise the required shape of $\Phi(x,y)$ is to consider the planar-type BH horizon surface, so the metric potential has to satisfy the elliptic wave equation \rf{rhoeqpl}.

\begin{figure}[h]
\centering
\includegraphics[width=9cm]{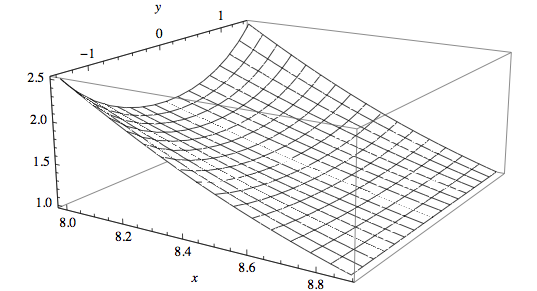}
\caption{Inverse of the true inhomogeneity distribution density on the horizon surface for the metric potential \rf{Phitr}.}
\label{figure:fig1}
\end{figure}

\ssk
By trials and errors method we found the following trial form of $\Phi(x,y)$, which falls within the above-mentioned criteria:
\be
\Phi(x,y)=\fr12 \sin \left(\fr1{\sqrt{2}}(x+iy)\right)+\fr12 \sin \left(\fr1{\sqrt{2}}(x-iy)\right)=\sin \left(\fr{x}{\sqrt{2}}\right)\cosh \left(\fr{y}{\sqrt{2}}\right).
\la{Phitr}
\ee
To satisfy the upper bound $\eta/s \lesssim 0.2$, the BH horizon should be restricted in $x,y$ directions to $x\in [-0.9+2\sqrt{2}\,\pi,2\sqrt{2}\,\pi]$, $y \in [-\sqrt{2},\sqrt{2}]$ (see Fig.\ref{figure:fig1}). 
Then, depending on the local value of the true inhomogeneity distribution density on the BH horizon surface $\exp(\Phi(x,y))$,\footnote{More precisely, depending on its inverse (cf. \rf{shear}).} one recovers the whole range of theoretical and experimental values $0.08 \le \eta/s \le 0.2$ with the entropy density $s$ measured in units of $s_0$. This example illustrates advantages of the developed approach, when having unfixed functions converts fitting to experimental data into a merely technical task.

\section{Summary and conclusions}

To summarise, we have obtained solutions to the Einstein--Maxwell dynamical system, which correspond to the static charged AdS black holes with inhomogeneity distribution function on the black hole horizon surface. The inhomogeneity of 2D horizon surface is encoded in the conformal factor entering the metric ansatz, which depends on the horizon coordinates and whose dynamics obeys the Liouville equation. That is why we have called such ingredient of the metric as the Liouville mode.

\ssk
Focusing on the AdS$_4$ space-time we have computed the charge diffusion coefficient and the DC conductivity on the horizon within the stretched horizon approach and have observed that: 
\begin{enumerate}
\item
The resulting Fick's laws describe diffusion in inhomogeneous strongly coupled dual media, which is natural to expect.
\item
The diffusion coefficient is exponentially suppressed, which may in principle violate the previously suggested KR universal bound for the diffusion constant \cite{Kovtun:2008kx}. In all possible cases the violation degree is proportional to the strength of the local inhomogeneity.
\end{enumerate}

\ssk
We have also calculated the charge diffusion and transport coefficients in the hydrodynamic limit of AdS/CFT correspondence and have realised that the KSS shear viscosity-per-entropy density universal bound \cite{Kovtun:2003wp, Kovtun:2004de}
is also exponentially suppressed. Hence, we have observed the violation of the KSS/KR universal bounds in backgrounds of charged black holes with planar/spherical/hyperbolic horizons within the standard Einstein--Maxwell setup. In all these cases we have observed that the violation of universal bounds depends on the explicit choice of the inhomogeneity distribution on the horizon and may be in general realised in local domains of its positivity. To show the relevance of the approach in situation when universal bounds hold we have given an example of the inhomogeneity distribution function, which preserves the KSS $\eta/s_0 \sim 0.08$ universal bound and fits all the range of experimentally measured at RHIC and at the LHC values of $\eta/s$ ratio in $s_0$ units. 

\ssk
The extension of the obtained solutions for RN black holes with constant curvature inhomogeneous horizons to higher-dimensional AdS spaces revealed the appearance of two different conductivities in 4D effective charged dual media, the corresponding ratio of which, within domains of preserving the KSS/KR bounds, possesses the same qualitative feature as that of previously found in 4D anisotropic strongly coupled plasma \cite{Rebhan:2011vd}. 

\ssk
Turning back to occurrence of the Liouville equation in condensed matter physics problems, we recall that the Liouville field theory naturally appears in CMP models related to diffusion processes in random media, or to the description of strongly coupled disordered media \cite{Kogan:1996wk,CD01,Ferrari:2009vi,Ferrari:2006ap}. We believe that our results open a new prospect in searching for holographic description of physical processes in disordered media at the strong coupling constant regime. We hope to report on progress in this and other directions in the future.

\bsk\bsk

\centerline{\bf Acknowledgements}
T.~M. is grateful to CERN Student Summer Programme for financial support and to CERN ATLAS group for kind hospitality during the course of this work. A.~N. acknowledges I.~V.~Pavlenko for valuable discussions on plasma physics.

\newpage

\bsk\bsk

\section*{Appendix A: Notation and conventions}

\addcontentsline{toc}{section}{Appendix A: Notation and conventions}

\def\theequation{A.\arabic{equation}}
\setcounter{equation}0

We use the mostly plus metric signature $(-,+,\dots,+)$ in $D$ dimensions. The coordinate system used in the paper is parameterised by coordinates $X^m=(t,r,x,y,\dots)$ where $t$ is the temporal coordinate, $r$ is the radial coordinate and the subset $(x,y,\dots)$ parameterises a $(D-2)$-dimensional space-like surface, which is called the horizon, $\mathcal{H}_{D-2}$. 

The metrics considered here correspond to $D$-dimensional Reissner--N\"ordstrom black holes  whose geometry is described by the space-time interval
\be
\mathrm{d}s^2=g_{mn} \,\mathrm{d}X^m\,\mathrm{d}X^n=-f(r)\mathrm{d}t^2+\fr{\mathrm{d}r^2}{f(r)}+r^2 \g_{ij}(r,X)\mathrm{d}{X}^i \mathrm{d}X^j.
\la{genint}
\ee
$\g_{ij}$ is the internal metric on the horizon surface. Together with \rf{genint} we use another representation of the space-time metric
\[
\mathrm{d}s^2=g_{tt}(r)\,\mathrm{d}t^2+g_{rr}(r) \,\mathrm{d}r^2+g_{ij}\, \mathrm{d}X^i \mathrm{d}X^j.
\]

Following \cite{Chamblin:1999tk, Chamblin:1999hg} it is convenient to introduce a constant $\w_D$ related to the volume of a $D-2$-dimensional horizon $\mathcal{H}_{D-2}$:
\be
\w_{D-2}=\fr{16 \pi G}{(D-2) V_{D-2}},\quad V_{D-2}=\int_{\mathcal{H}_{D-2}}\,\sqrt{\g}\,\mathrm{d}x^1\dots \mathrm{d}x^{D-2},\qquad \g\equiv \det \g_{ij}.
\la{wdef}
\ee

\bsk
\section*{Appendix B: Real solutions to the elliptic wave/Liouville equation}

\addcontentsline{toc}{section}{Appendix B: Real solutions to the elliptic wave/Liouville equation}

\def\theequation{B.\arabic{equation}}
\setcounter{equation}0

Solution to the elliptic wave equation 
\[
\fr{\pa^2 \Phi}{\pa x^2}+\fr{\pa^2 \Phi}{\pa y^2}=0
\]
is well known
\be
\Phi(x,y)={f} (x+iy)+{g}(x-iy)\equiv f(u)+g(\bar{u}).
\la{rhoeqplsol}
\ee
Here we have introduced the complex ``light-cone'' variables
\be
u=\fr1{\sqrt{2}}(x+iy),\qquad \bar{u}=\fr1{\sqrt{2}}(x-iy);
\la{ubarudef}
\ee
$f(u),g(\bar{u})$ are arbitrary complex functions. 
It is also convenient to introduce the Wirtinger derivatives (see \cite{Henrici74})
\be
\pa_u=\fr1{\sqrt{2}}(\pa_x-i\pa_y),\qquad \pa_{\bar{u}}=\fr1{\sqrt{2}}(\pa_x+i\pa_y),
\la{paubaru}
\ee
which act as $\pa_u u=1$, $\pa_{\bar{u}}\bar{u}=1$. Physically motivated requirement of a real valued metric potential, with $\mathrm{Im}\, \Phi(x,y)=0$, restricts the realisation of $f(u)$ and $g(\bar{u})$ in terms of elementary and special functions. For example,
\be
f(u)=\fr12 (x+iy)^2, \qquad g(\bar{u})=\fr12 (x-iy)^2\,;
\la{solspl1}
\ee
or
\be
f(u)=\sin \left[\fr1{\sqrt{2}}(x+iy)\right],\qquad g(\bar{u})=\sin \left[\fr1{\sqrt{2}} (x-iy) \right].
\la{solspl2}
\ee

\bsk
The elliptic Liouville equation
\be
\fr{\pa^2 \Phi}{\pa x^2}+\fr{\pa^2 \Phi}{\pa y^2}+2K\mathrm{e}^{\Phi(x,y)}=0
\la{elLeq}
\ee
in complex ``light-cone'' coordinates is simplified to
\be
\pa_u \pa_{\bar{u}} \Phi(u,\bar{u})=-K \mathrm{e}^{\Phi(u,\bar{u})}.
\la{elubuL}
\ee
For $K>0$ the most general real solution to the elliptic Liouville equation can be found in \cite{Crowdy97}. As an illustrative example, we give one of solutions by Crowdy, related to the original Liouville solution:
\be
\Phi(u,\bar{u})=-2\ln \left[\sqrt{\fr{K}{2}} (f(u)\bar{f}(\bar{u})+1)\right]+\ln \left[f'(u)\bar{f}'(\bar{u}) \right],
\la{Cr}
\ee
where $f(u)=F_1(x,y)+i F_2(x,y)$ with arbitrary real functions $F_{1,2}$. The case of $K<0$ is more subtle; we give three solutions by Popov \cite{Popov93}, mentioned in \cite{Crowdy97}:
\be
\Phi=\ln \left[\fr{1}{F(v)}\left((\pa_x v)^2+(\pa_y v)^2 \right)\right],\qquad 
F(v)=\left\{
v^2,
\sin^2 v,
\sinh^2 v
\right\},\qquad K=-1,
\la{Po}
\ee
where $v(x,y)$ is the real part of a general analytic function $f(x+iy)$.

\ssk
\ssk


\begin{thebibliography}{99}

\bibitem{Kovtun:2003wp} 
  P.~Kovtun, D.~T.~Son, A.~O.~Starinets,
  Holography and hydrodynamics: Diffusion on stretched horizons.
  JHEP {\bf 0310}, 064 (2003).
  arXiv:hep-th/0309213

\bibitem{Kovtun:2004de} 
  P.~Kovtun, D.~T.~Son, A.~O.~Starinets,
  Viscosity in strongly interacting quantum field theories from black hole physics.
  Phys.\ Rev.\ Lett.\  {\bf 94}, 111601 (2005).
  arXiv:hep-th/0405231

\bibitem{Kovtun:2008kx} 
  P.~Kovtun, A.~Ritz,
  Universal conductivity and central charges.
  Phys.\ Rev.\ D {\bf 78}, 066009 (2008).
  arXiv:0806.0110 [hep-th]


\bibitem{Ritz:2010zza} 
  A.~Ritz,
  Probing universality in AdS/CFT.
  Int.\ J.\ Mod.\ Phys.\ A {\bf 25}, 433 (2010)


\bibitem{Horowitz:2008bn} 
  G.~T.~Horowitz, M.~M.~Roberts,
  Holographic superconductors with various condensates.
  Phys.\ Rev.\ D {\bf 78}, 126008 (2008).
  arXiv:0810.1077 [hep-th]


\bibitem{Iqbal:2008by} 
  N.~Iqbal, H.~Liu,
  Universality of the hydrodynamic limit in AdS/CFT and the membrane paradigm.
  Phys.\ Rev.\ D {\bf 79}, 025023 (2009).
  arXiv:0809.3808 [hep-th]


\bibitem{Brigante:2007nu} 
  M.~Brigante, H.~Liu, R.~C.~Myers, S.~Shenker, S.~Yaida,
  Viscosity bound violation in higher derivative gravity.
  Phys.\ Rev.\ D {\bf 77}, 126006 (2008).
  arXiv:0712.0805 [hep-th]


\bibitem{Ritz:2008kh} 
  A.~Ritz, J.~Ward,
  Weyl corrections to holographic conductivity.
  Phys.\ Rev.\ D {\bf 79}, 066003 (2009).
  arXiv:0811.4195 [hep-th]


\bibitem{Wu:2010vr} 
  J.~-P.~Wu, Y.~Cao, X.~-M.~Kuang, W.~-J.~Li,
  The 3+1 holographic superconductor with Weyl corrections.
  Phys.\ Lett.\ B {\bf 697}, 153 (2011).
  arXiv:1010.1929 [hep-th]


\bibitem{Erdmenger:2010xm} 
  J.~Erdmenger, P.~Kerner, H.~Zeller,
  Non-universal shear viscosity from Einstein gravity.
  Phys.\ Lett.\ B {\bf 699}, 301 (2011).
  arXiv:1011.5912 [hep-th]


\bibitem{Mateos:2011ix} 
  D.~Mateos, D.~Trancanelli,
  The anisotropic N=4 super Yang-Mills plasma and its instabilities.
  Phys.\ Rev.\ Lett.\  {\bf 107}, 101601 (2011).
  arXiv:1105.3472 [hep-th]

\bibitem{Mateos:2011tv} 
  D.~Mateos, D.~Trancanelli,
  Thermodynamics and instabilities of a strongly coupled anisotropic plasma.
  JHEP {\bf 1107}, 054 (2011).
  arXiv:1106.1637 [hep-th]


\bibitem{Rebhan:2011vd} 
  A.~Rebhan, D.~Steineder,
  Violation of the holographic viscosity bound in a strongly coupled anisotropic plasma.
  Phys.\ Rev.\ Lett.\  {\bf 108}, 021601 (2012).
  arXiv:1110.6825 [hep-th]


\bibitem{Critelli:2014kra} 
  R.~Critelli, S.~I.~Finazzo, M.~Zaniboni, J.~Noronha,
  Anisotropic shear viscosity of a strongly coupled non-Abelian plasma from magnetic branes.
Phys.\ Rev.\ D {\bf 90}, 066006 (2014). arXiv:1406.6019 [hep-th]


\bibitem{Lemos:1994fn} 
  J.~P.~S.~Lemos,
  Two-dimensional black holes and planar general relativity.
  Class.\ Quantum Gravity  {\bf 12}, 1081 (1995).
  arXiv:gr-qc/9407024

\bibitem{Vanzo:1997gw} 
  L.~Vanzo,
  Black holes with unusual topology.
  Phys.\ Rev.\ D {\bf 56}, 6475 (1997).
  arXiv:gr-qc/9705004

\bibitem{SpivakV4}
M.~Spivak, A Comprehensive Introduction to Differential Geometry, vol. 4, 3rd edn. (Publish or Perish, Houston, 1999)

\bibitem{Parikh:1997ma} 
  M.~Parikh, F.~Wilczek,
  An action for black hole membranes.
  Phys.\ Rev.\ D {\bf 58}, 064011 (1998).
  arXiv:gr-qc/9712077

\bibitem{Damour:1978cg} 
  T.~Damour,
  Black hole eddy currents.
  Phys.\ Rev.\ D {\bf 18}, 3598 (1978)

\bibitem{Blandford:1977ds} 
  R.~D.~Blandford, R.~L.~Znajek,
  Electromagnetic extractions of energy from Kerr black holes.
  Mon.\ Not.\ Roy.\ Astron.\ Soc.\  {\bf 179}, 433 (1977)
  
\bibitem{Znajek:1978} 
  R.~L.~Znajek, The electric and magnetic conductivity of a Kerr hole.
  Mon.\ Not.\ Roy.\ Astron.\ Soc.\  {\bf 185}, 833 (1978)

\bibitem{MPbook}
K.~S.~Thorne, D.~A.~MacDonald, R.~H.~Price (eds.), Black Holes: The Membrane Paradigm (Yale University Press, New Haven, 1986)

\bibitem{BDbook}
N.~D.~Birrell, P.~C.~W.~Davies, Quantum Fields in Curved Space (Cambridge University Press, Cambridge, 1982)

\bibitem{Son:2002sd} 
  D.~T.~Son, A.~O.~Starinets,
  Minkowski space correlators in AdS / CFT correspondence: recipe and applications.
  JHEP {\bf 0209}, 042 (2002).
  arXiv:hep-th/0205051



\bibitem{Policastro:2002se} 
  G.~Policastro, D.~T.~Son, A.~O.~Starinets,
  From AdS / CFT correspondence to hydrodynamics.
  JHEP {\bf 0209}, 043 (2002).
  arXiv:hep-th/0205052




\bibitem{Maldacena:1997re} 
  J.~M.~Maldacena,
  The large N limit of superconformal field theories and supergravity.
  Int.\ J.\ Theor.\ Phys.\  {\bf 38}, 1113 (1999)
  
\bibitem{Maldacena:1998}
  J.~M.~Maldacena,
  Adv.\ Theor.\ Math.\ Phys.\  {\bf 2}, 231 (1998).
  arXiv:hep-th/9711200


\bibitem{Gubser:1998bc} 
  S.~S.~Gubser, I.~R.~Klebanov, A.~M.~Polyakov,
  Gauge theory correlators from noncritical string theory.
  Phys.\ Lett.\ B {\bf 428}, 105 (1998).
  arXiv:hep-th/9802109

\bibitem{Witten:1998qj} 
  E.~Witten,
  Anti-de Sitter space and holography.
  Adv.\ Theor.\ Math.\ Phys.\  {\bf 2}, 253 (1998).
  arXiv:hep-th/9802150



\bibitem{Son:2006em} 
  D.~T.~Son, A.~O.~Starinets,
  Hydrodynamics of r-charged black holes.
  JHEP {\bf 0603}, 052 (2006).
  arXiv:hep-th/0601157


\bibitem{Natsuume:2010ky} 
  M.~Natsuume, M.~Ohta,
  The shear viscosity of holographic superfluids.
  Prog.\ Theor.\ Phys.\  {\bf 124}, 931 (2010).
  arXiv:1008.4142 [hep-th]


\bibitem{Herzog:2002fn} 
  C.~P.~Herzog,
  The hydrodynamics of M theory.
  JHEP {\bf 0212}, 026 (2002).
  arXiv:hep-th/0210126



\bibitem{Crowdy97}
D.~G.~Crowdy, General solutions to the 2D Liouville equation. Int. J. Eng. Sci. {\bf 35}, 141 (1997)


\bibitem{Popov93}
A.~G.~Popov, Exact formulae of constructing solutions to the Liouville equation by use of solutions to the Laplace equation. Dokl. Akad. Nauk 333 {\bf 4}, 440--441 (1993) (in Russian)

\bibitem{LAGG91}
L. Alvarez-Gaume, C. Gomez, Topics in Liouville Theory. Lectures at the Spring School on String Theory and Quantum Gravity, Trieste, Italy, Apr 15--23, 1991, Published in Trieste
Spring School 1991, pp. 142--177

\bibitem{abdalla01}
E.~Abdalla, M.~C.~B. Abdalla, K.~D.~Rothe, Non-Perturbative Methods in 2 Dimensional Quantum Field Theory, 2nd edn. (World Scientific, Singapore, 2001)



\bibitem{Kogan:1996wk} 
  I.~I.~Kogan, C.~Mudry, A.~M.~Tsvelik,
  The Liouville theory as a model for prelocalized states in disordered conductors.
  Phys.\ Rev.\ Lett.\  {\bf 77}, 707 (1996).
  arXiv:cond-mat/9602163

\bibitem{CD01}
D.~Carpentier and P.~Le Doussal, Glass transition of a particle in a random potential, front selection in nonlinear renormalization group, and entropic phenomena in Liouville and sinh-Gordon models. Phys. Rev. E {\bf 63}, 026110 (2001). arXiv:cond-mat/0003281 [Erratum-ibid. E {\bf 73}, 019910 (2006)]


\bibitem{Ferrari:2006ap} 
  F.~Ferrari, J.~Paturej,
  On a relation between Liouville field theory and the Brownian motion of particles.
  Phys.\ Lett.\ B {\bf 664}, 123 (2008).
  arXiv:math-ph/0609058


\bibitem{Ferrari:2009vi} 
  F.~Ferrari, J.~Paturej,
  Diffusion of Brownian particles and Liouville field theory.
  Acta Phys.\ Polon.\ B {\bf 40}, 1383 (2009).
  arXiv:0901.1234 [hep-th]

\bibitem{HBA02}
S.~Havlin, D.~Ben-Avraham, Diffusion in disordered media. Adv. Phys. {\bf 51}, 187--292 (2002)


\bibitem{Dvali:2012en} 
  G.~Dvali, C.~Gomez,
  Black holes as critical point of quantum phase transition.
  Eur.\ Phys.\ J.\ C {\bf 74}, 2752 (2014).
  arXiv:1207.4059 [hep-th]


\bibitem{Germani:2013sra} 
  C.~Germani,
  On the many saddle points description of quantum black holes.
  Phys.\ Lett.\ B {\bf 733}, 93 (2014).
  arXiv:1307.6238 [hep-th]


\bibitem{Mamo:2012sy} 
  K.~A.~Mamo,
  Holographic RG flow of the shear viscosity to entropy density ratio in strongly coupled anisotropic plasma.
  JHEP {\bf 1210}, 070 (2012).
  arXiv:1205.1797 [hep-th]


\bibitem{Gale:2012rq} 
  C.~Gale, S.~Jeon, B.~Schenke, P.~Tribedy, R.~Venugopalan,
  Event-by-event anisotropic flow in heavy-ion collisions from combined Yang-Mills and viscous fluid dynamics.
  Phys.\ Rev.\ Lett.\  {\bf 110}, 012302 (2013).
  arXiv:1209.6330 [nucl-th]

\bibitem{Song:2012ua} 
  H.~Song,
  QGP viscosity at RHIC and the LHC --- a 2012 status report.
  Nucl.\ Phys.\ A {\bf 904--905}, 114c (2013).
  arXiv:1210.5778 [nucl-th]


\bibitem{Chamblin:1999tk} 
  A.~Chamblin, R.~Emparan, C.~V.~Johnson and R.~C.~Myers,
  Charged AdS black holes and catastrophic holography.
  Phys.\ Rev.\ D {\bf 60}, 064018 (1999).
  arXiv:hep-th/9902170


\bibitem{Chamblin:1999hg} 
  A.~Chamblin, R.~Emparan, C.~V.~Johnson, R.~C.~Myers,
  Holography, thermodynamics and fluctuations of charged AdS black holes.
  Phys.\ Rev.\ D {\bf 60}, 104026 (1999).
  arXiv:hep-th/9904197


\bibitem{Henrici74}
P.~Henrici, Applied and Computational Complex Analysis: Vol. 3  Discrete Fourier Analysis, Cauchy Integrals, Construction of Conformal Maps, Univalent Functions, (Wiley, New York, 1974)







\end{thebibliography}
\end{document}